# LM-PROTAC: a language model-driven PROTAC generation pipeline with dual constraints of structure and property

Jinsong Shao[1#], Qineng Gong[2#], Zeyu Yin[1], Yu Chen[1], Yajie Hao[1], Lei Zhang[3], Linlin Jiang[3], Min Yao[4], Jinlong Li[3], Fubo Wang[5,6*], Li Wang[7*].

1. School of Information Science and Technology, Nantong University, Nantong 226001
2. Medical Research Center, Affiliated Hospital 2 of Nantong University and First People's Hospital of Nantong City, Nantong 226001, China
3. School of Pharmacy, Nantong University, Nantong 226001, China
4. School of Medical, Nantong University, Nantong 226001, China
5. Center for Genomic and Personalized Medicine, Guangxi key Laboratory for Genomic and Personalized Medicine
6. Guangxi Collaborative Innovation Center for Genomic and Personalized Medicine, Guangxi Medical University, Nanning 226001, China
7. Research Center for Intelligence Information Technology, Nantong University, Nantong 226001, China

* Corresponding authors
# These authors contributed equally to this work.

## Abstract

The imperfect modeling of ternary complexes has limited the application of computer-aided drug discovery tools in PROTAC research and development. In this study, a language model for PROTAC molecule design pipeline named LM-PROTAC was developed, which stands for language model-driven Proteolysis Targeting Chimera, by embedding a transformer-based generative model with dual constraints on structure and properties. This study started with the idea of segmentation and representation of molecules and protein. Firstly, a language model-driven affinity model for protein compounds to screen molecular fragments with high affinity for the target protein. Secondly, structural and physicochemical properties of these fragments were constrained during the generation process to meet specific scenario requirements. Finally, a two-round screening was performed on the preliminary generated molecules using a multidimensional property prediction model. This process identified a batch of

PROTAC molecules capable of degrading disease-relevant target proteins. These molecules were subsequently validated through in vitro experiments, thus providing a complete solution for language model-driven PROTAC drug generation. Taking Wnt3a, a key tumor-related target, as a POI of degradation, the LM-PROTAC pipeline successfully generated effective PROTAC molecules. The molecular distribution experiments demonstrated the high similarity of the generated molecules to the original dataset, validating the generative model's effectiveness in accurately defining chemical space. Molecular dynamics simulations confirmed the stable interactions between the PROTAC molecules and target proteins, while protein degradation experiments verified the efficacy of the generated PROTAC molecules in degrading target proteins. The entire LM-PROTAC pipeline is reusable and can generate degraders for other target proteins within 50 days, significantly improving the efficiency of drug discovery for undruggable targets.

**Introduction**

PROTAC hijacks the activity of E3 ubiquitin ligase to ubiquitinate the POI, leading to its degradation by the 26S proteasome and mediating the degradation of the POI. This hijacking mechanism has been employed to degrade various types of disease-related POIs[1]. Over the past two decades, there has been continuous effort to target the Ubiquitin-proteasome System (UPS) for therapeutic purposes. The approval of proteasome inhibitors by the U.S. Food and Drug Administration has demonstrated the pharmacological potential of UPS, prompting further research and expansion into manipulating this pathway for disease treatment. Unlike traditional small molecule inhibition principles, UPS can be artificially intervened through a Target based degradation strategy to selectively target and degrade specific proteins. Currently, many approaches are being developed by researchers, including protein-targeting chimeras, AdPROMs[2], biological PROTAC[3], molecular glues, and selective estrogen receptor downregulators[4, 5].

In the early development of PROTAC, the first successful case of protein degradation mediated by PROTAC was achieved using pPROTAC[6]. However, due to issues such as the instability and poor cell permeability of pPROTAC itself as a peptide molecule, the research focus gradually shifted towards small molecule PROTAC[7-9]. As a result, the study of small molecule PROTAC in the field has gained more recognition among industry researchers. AI is experiencing robust growth. Among various AI methods, generation models have gained considerable attention in recent years. In addition, NLP techniques offer new possibilities for the design of PROTAC molecules, particularly when targeting proteins that lack well-defined binding pockets. Traditional small-molecule drug design relies on the binding pockets of target proteins, while NLP techniques enable the identification of various binding sites by fragmenting molecules. This technique has the potential to facilitate the design of efficient PROTAC molecules, potentially expanding the application of PROTAC. Inspired by these successful developments, researchers are now applying generation model technology to de novo drug design, considered to be the origin of drug discovery. From this perspective, various models such as Recurrent Neural Networks, Autoencoders, Generative Adversarial Networks, Transformers, and Hybrid Models with Reinforcement Learning have demonstrated exceptional capabilities in various molecular generation tasks. Consequently, applying these de novo drug design approaches to the scenario of PROTAC drug generation has become an important application in the specialized field.

**Molecular and protein encoding**

Molecular representation is a crucial task in the molecular generation workflow. Researchers have constructed models for accurate molecular representation from multiple dimensions and perspectives, including 1D[10-15], 2D[16-19], 3D[20-23], and images[24-26], and validated them through experiments. However, most attention has been given to molecular representation methods based on atomic and bond structures, overlooking the impact of interactions between molecular structural fragments on molecular properties. Therefore, from the perspective of a language model, molecules are divided into combinations of multiple fragments, focusing on the interaction relationships among these token fragments to identify those fragments that have a significant impact on molecular properties.

Molecular and protein sequences are analogous to full sentences in natural language. Initially, molecules and proteins are segmented into fragments called *S-mol* and *S-pro*, representing Segment molecular and Segment protein, respectively. This segmentation approach is inspired by the concept of "Segment" in NLP, simplifying complex molecular and protein sequences into relatively complete functional fragments. Subsequently, S-mol and *S-pro* are further divided into smaller units called *T-mol* and *T-pro*, derived from Token molecular and Token protein.

## Structure constraint

Molecules are composed of *S-mol*, and *S-mol* represent the fundamental chemical structures that exhibit various properties of the molecule. Research on molecular *S-mol*s allows for the understanding of interactions between internal or intermolecular local *S-mol*s. Interactions between molecules arise from the collective interactions among molecular *S-mol*s, making high-affinity *S-mol* a crucial source of molecular affinity. Constructing a molecule based on the generation of high-affinity *S-mol* has the potential to yield molecules with high affinity. Similar to natural language, molecular encoding as a chemical language relies on segmentation that adheres to specific chemical logic, which is a crucial prerequisite for a deeper understanding of molecular encoding. Therefore, successful segmentation methods applied to natural language can also be applied to molecular encoding segmentation.

Reinforcement learning is a target-oriented machine learning approach that takes environmental feedback as input and adapts to the environment. Its main idea is to find the optimal behavioral strategy by interacting with the environment through trial and error, mimicking the fundamental way humans or animals learn. The core principle of reinforcement learning is to learn a series of actions that guide the model to achieve its goal or maximize its objective function. If an action by the agent leads to a positive reward from the environment, i.e., a reinforced signal, the tendency of the agent's subsequent actions will be strengthened. Otherwise, the inclination of the agent to produce such actions will weaken. This is consistent with the principles of classical conditioning in physiology. Structural constraints in molecular generation are achieved by reinforcing molecules that are generated with structures closer to the desired constrained structure through repeated rewards in reinforcement learning.

Molecular structures are constrained through the application of reinforcement learning. To preserve the structural features of highly affine molecular *S-mol* in the generated molecules, reinforcement learning is employed by imposing constraints on the similarity between the generated molecules and target *S-pro*. The model rewards the agent with positive reinforcement each time a newly generated molecule is structurally closer to the target *S-pro* and imposes penalties when the generated molecule deviates from the target *S-pro*. The use of reinforcement learning avoids the low degrees of freedom approach of directly connecting target *S-pro*s, thus preventing limitations on the model's ability to generate entirely new molecules. A structure-constrained molecular generation strategy involves restricting the output molecules to contain a specific skeleton or *S-mol*. Langevin et al. and Li et al. established generative models that output drug molecules with specific skeletons[27, 28]. These skeletons are often extracted from existing drugs with favorable biological properties. Several researchers also developed skeleton-based generative models, learning to generate molecules with specific *S-mol*s[29-32]. However, structure-constrained molecular generation models often produce a large number of repetitive structures and molecules, limiting the model's freedom by constraining the primary structure of the molecular skeleton. This results in the generation of numerous structurally similar molecules for the same drug, thereby reducing the model's learning and generative capabilities for new drugs. Therefore, considering multiple possible high-affinity *S-pro*s in generative models and using them as a basis for molecular generation, along with the application of reinforcement learning, enables the generated molecules to retain the structural features of the target *S-pro*s while introducing variability based on fixed structural characteristics. This enhances the model's ability to generate diverse molecules.

**Physicochemical Property Constraint**

Currently, a large number of drug generation models have been developed by researchers, such as VAE[33-38], GF[39, 40], AAE[41], and others. However, during the development of these models, a significant issue has gradually emerged. Most models only focus on the biological activity of drugs to targets, with a few models considering one or a few other drug properties, such as drug concentration, solubility, etc. When a compound is gradually classified as a drug, a large number of drug properties need to be considered. These drug properties contribute differently to the compound's drugability. In many generative models, important drug properties such as water solubility and lipid solubility are not taken into account. These crucial drug properties are often incorporated only in the subsequent screening of potential drugs, resulting in additional costs during both the generation process and the later dry and wet experiments for property screening. Therefore, it is necessary to consider some important properties as constraints during the drug generation process.

There are multiple reasons why people might be interested in discovering new molecules. To integrate generative models into molecular design, it is essential to define these various applications as specific problem statements. For example, molecules that possess a particular property X, may be discovered while certain constraints Y, are met. The generated molecules are enhanced by adding property constraints, ensuring that the compounds are chemically valid and exhibit specific desirable properties, such as good solubility, low toxicity, or high potency. Since it is impractical to experimentally validate each generated compound, it becomes necessary to train a property predictor to assess compound properties, also known as a QSAR model. The property predictor is trained on a separate molecular dataset labeled with their properties (e.g., $IC_{50}/EC_{50}$ for potency). When the training is completed, the property predictor is used to estimate whether the generated molecules satisfy the given constraint conditions. In this way, the generative model learns to generate compounds predicted by the property predictor to meet the constraints. This task is often considered a discrete optimization problem and can be addressed using reinforcement learning, Bayesian optimization, or genetic algorithms. In reinforcement learning, a model is trained to maximize the expected reward based on the property predictor's output. Additionally, Bayesian optimization methods can transform discrete optimization problems into continuous optimization problems by learning continuous embeddings of molecules through methods such as training a VAE to map discrete molecules into a continuous embedding space. Another neural network is then trained to predict the chemical properties of the original molecules from their continuous embedding vectors. Bayesian optimization is then applied in the continuous embedding space to find an embedding with the best correlation property scores. This optimal embedding is decoded by a decoder network into a discrete molecule. Genetic algorithms solve discrete optimization problems by searching for favorable compounds through molecular mutations. Genetic algorithms consist of a set of mutation rules and a fitness function. The fitness function is a weighted interpolation of predicted property scores and penalty scores for long-lived molecules. Additional penalty terms encourage the model to explore a diverse set of molecules. New compounds are derived by applying mutation rules to existing molecules. In each iteration, molecules with lower fitness scores are removed. Nigam et al. applied genetic algorithms to design molecules with high logP scores, where the fitness function was parameterized as a neural network[42]. It is worth noting that while many studies use logP as a convenient metric for method development, it is an artificial task that cannot be directly tied to any practical application[42].

**Molecular generation by language model**

Molecular generation can be viewed from the perspective of natural language processing as the generation of language sequences. The Transformer model has demonstrated state-of-the-art performance in natural language processing[43, 44], and it has recently found applications in the field of drug generation. The original version of the Transformer consists of an encoder and a decoder, with a key feature being the attention mechanism that can capture long-range dependencies in sequences. Hybrid models, which combine deep generative models with reinforcement learning, have been applied to generate molecules from scratch biased towards desired properties[38]. The Transformer, being a sequence-based model, exhibits characteristics that are well-suited for molecular representation, particularly in 1D sequence encoding similar to SMILES. The multi-head attention mechanism of Transformer, which focuses on long-range dependencies, aligns with the nature of molecular sequences where distant segments produce remote correlations. Introducing the Transformer in the representation of compounds allows it to capture these distant correlations, considering the interactions between segments that are far apart in the sequence. The unified interactions among these local segments constitute the interactions between molecules. These segments, representing local information in the molecule, serve as crucial carriers of information for achieving precise molecular representation. A rational splitting logic enhances the accuracy of functional information division and improves the reliability and stability for molecular generation based on *S-mol* structure constraints. The C-Transformer model is a conditional language generation model that originated from Salesforce's work on targeted writing[45]. Hou et al. introduced the Conditional Transformer into the field of molecular generation. By incorporating a conditional

Transformer model, they generated molecules that meet specified property requirements, imposing constraints on molecular properties[46].

Compared to generated molecules by C-Transformer, traditional approaches, such as the Rosetta method, use physical energy functions and high-resolution models for protein structure prediction and molecular design, providing high-accuracy results. This is particularly important for designing and optimizing protein-ligand interactions. Rosetta is also versatile, as it can be used not only for protein-ligand docking but also for protein design, protein structure prediction, free energy calculations, and small molecular design, among other applications. Additionally, by simulating folding and energy minimization, Rosetta can identify the most stable molecular conformations, thereby enhancing the affinity and stability of molecules. However, the computational process of Rosetta is complex and time-consuming, requiring significant computational resources and time, especially when dealing with large molecular systems. The accuracy of the predictions also depends on the quality of the input data; any inaccuracies in the input can affect the results. For large molecular systems or complex setups, the computational difficulty and time costs of Rosetta increase significantly[47-49].

Additionally, molecular docking methods can rapidly screen a large number of molecules to identify potential high-affinity ligands, making them suitable for initial screening[50]. Compared to full physical simulations, docking methods are more straightforward and faster, allowing for the processing of numerous molecules in a short time. They have been widely applied in drug discovery, demonstrating their effectiveness. However, docking methods often simplify the physical and chemical processes of the system, which may lead to insufficient predictive accuracy. Furthermore, due to the simplified models, docking results may contain a significant number of false positives that require further experimental validation. The results of docking are highly dependent on the initial conformations, which may cause some important binding modes to be overlooked[51, 52].

In the field of drug design, traditional molecular generation and screening methods, such as virtual enumeration and scoring methods, have become foundational tools in drug discovery. These methods help researchers identify potential drug molecules by exploring a vast number of possibilities in chemical space. However, as the complexity of drug design and the demands for precision increase, traditional methods have revealed significant limitations in terms of their ability to accurately predict molecular properties, assess synthetic feasibility, and handle the vast diversity of chemical structures. To overcome these challenges, researchers have begun to explore more intelligent and efficient molecular generation methods that leverage advanced algorithms and machine learning techniques to better address the evolving needs of modern drug discovery.

Virtual enumeration is a method that generates a large number of candidate molecules by combining predefined chemical *S-mol*. The main advantage of this approach is its ability to systematically cover a wide chemical space, resulting in millions of candidate molecules. However, a significant drawback of virtual enumeration is that the sheer number of generated molecules is overwhelming, with most lacking ideal bioactivity. This method relies on subsequent high-throughput screening techniques to identify a few potential candidate molecules, but this process consumes substantial computational resources and may lead to inefficient screening. Additionally, virtual enumeration typically does not take into account the specific physicochemical properties of the molecules and the requirements of the target, resulting in a variable quality of generated molecules and further complicating the screening process[53].

Scoring is a commonly used subsequent step in virtual screening to evaluate the binding affinity and pharmacological properties of candidate molecules with target proteins. The basis of scoring methods typically lies in molecular docking models or machine learning-based predictive models, which score molecules by calculating the interaction energy between them and the target protein. The advantage of scoring methods is that they provide researchers with a quantitative standard to help them filter out molecules that may possess high bioactivity. However, the effectiveness of scoring methods is highly dependent on the initial quality of the generated molecules. If the quality of the candidate molecules is low, it is difficult to identify truly effective drug molecules, even if the scoring model is very accurate. Since the generation and scoring processes are separate steps, scoring methods often fail to fully utilize the information obtained during molecular generation, leading to the possibility that the final selected molecules are not optimal[54, 55].

Language model-driven molecular generation methods may rapidly produce a large number of potentially bioactive molecules, greatly improving design efficiency. These models are efficient and flexible, capable of quickly adapting to different design tasks through training data, generating diverse molecular structures and exploring new chemical spaces that traditional methods find difficult to discover. Additionally, language models can generate molecules with novel structures and potential bioactivity, providing more options and possibilities for drug discovery. However, the molecules generated by these models may lack rigorous validation of their physical and chemical foundations, necessitating further experimental verification to ensure their actual effectiveness. The performance of the models largely depends on the quality and diversity of the training data; if the training data is insufficient or of low quality, the effectiveness of the generated molecules will also be impacted. The generated molecules need further optimization and screening to yield actual high-affinity candidates, which may require combining with other methods for subsequent optimization.

In this work, important functional proteins were targeted involved in disease progression. The problem was approached through the lens of language generation models, with reinforcement learning applied to constrain the structure of generated molecules. Additionally, the C-Transformer model was integrated to control the physicochemical properties of generated molecules. By combining the constraints on

*S-mol* structure and physicochemical properties, a comprehensive solution was developed for the targeted generation of PROTAC small molecule drugs aimed at specific protease targets. The feasibility of this approach was validated by using Wnt3a, an important early-stage target in liver cancer, as a case study.

## Methods and materials

### Dataset

The ZINC dataset was developed to bridge the gap between cheminformatics and biology. The ZINC database team developed a suite of ligand annotation, purchasability, target, and biology association tools, incorporated into ZINC and meant for investigators who are not computer specialists. The new version incorporates over 120 million purchasable drug-like compounds. ZINC links purchasable compounds with high-value compounds such as metabolites, drugs, natural products, and annotated compounds from the literature. Compounds can be accessed based on the genes annotated to them and the primary and secondary target classes to which these genes belong. It provides new analysis tools that are user-friendly for non-experts but have almost no limitations for experts[56]. ZINC is freely available at the following website: http://zinc15.docking.org.

BindingDB is a publicly accessible database released by the laboratory of Michael K. Gilson at the University of California, San Diego. It primarily collects non-covalent binding affinity data between drug target proteins and drug-like small molecules[57]. Researchers can access non-covalent binding data for relevant molecules, thereby facilitating drug development and the construction of binding prediction models. As of December 31, 2023, BindingDB's patent dataset includes, 6,765 patents, 1,059,214 binding measurements, 505,009 compounds, 2,578 target proteins, 9,728 assays.

The Davis dataset is a publicly available dataset primarily used for drug discovery and chemical biology research[58]. It includes binding affinity data between drugs and proteins. Researchers selected 68 drug compounds and 379 protein targets, using drug-protein pairs with binding affinities less than 30 units as positive samples.

The Biosnap dataset is another publicly available dataset focused on bioinformatics and chemical biology research, specifically targeting the prediction of interactions between compounds and proteins. This dataset contains 13,741 compound-protein interaction pairs, involving 4,510 different drugs and 2,181 protein targets.

The DUD-E database is a commonly used benchmark in structure-based virtual screening for evaluating the performance of various methods[59]. It includes 22,886 active ligands and their affinities against 102 targets.

All datasets are subjected to experiments using five-fold cross-validation. The datasets are divided into 80% training sets and 20% testing sets to ensure the stability and generalization ability of the model's performance. This partitioning method allows for the effective utilization of the information in the datasets during model training, ensuring consistent and reliable performance across different datasets.

## Solutions for PROTAC small molecule drug generation models

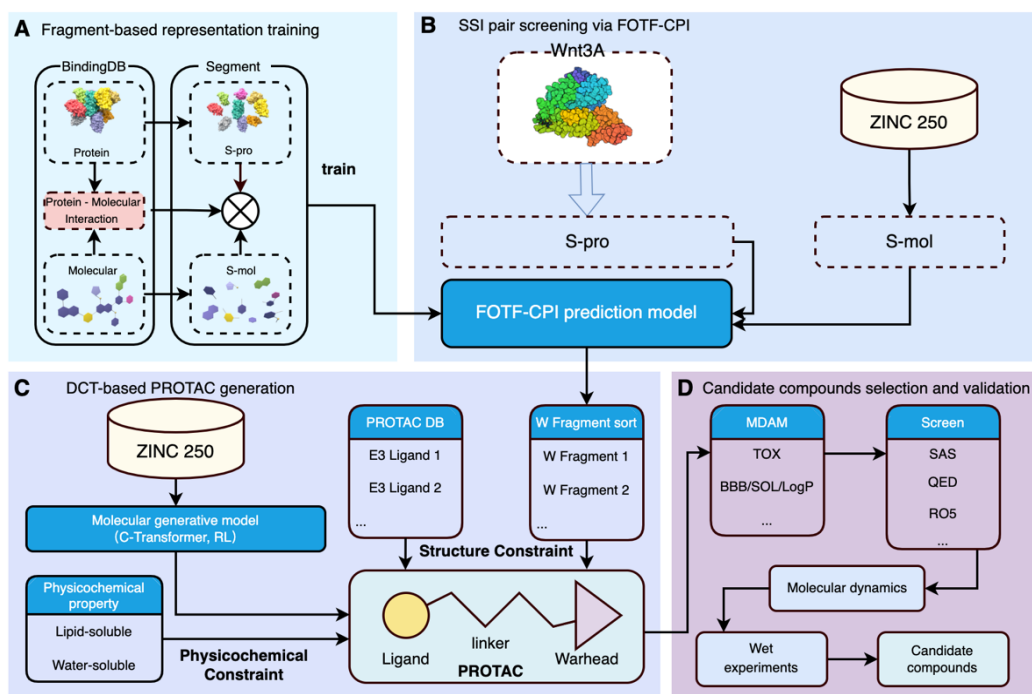

**Figure 1**. The main workflow of the LM-PROTAC pipeline

Figure 1 depicts the main workflow of the LM-PROTAC pipeline. Initially, molecular and protein representations are obtained by segmenting molecules and proteins from the dataset, which are used to train the FOTF-CPI model[60], as shown in Figure 1A. FOTF-CPI is a model designed for calculating the affinity between *S-mol*s and *S-pro*s. For the target protein Wnt3a, the model is employed to screen S-pro-S-mol Interaction Pairs, referred to as SSI pairs, with high affinity from the ZINC250 dataset, by calculating the global and local interaction relationships between *T-mol* and *T-pro*, as shown in Figure 1B. Figure 1C illustrates the construction of a PROTAC molecule generation model called DCT based on *S-mol* structure and physicochemical property constraints. This model, relying on C-Transformer and Reinforcement Learning, generates PROTAC molecules with high target affinity and specific attributes. The generated molecules meeting the requirements undergo molecular property filtering using a MDAM for further specific drug property requirements, resulting in potential candidate PROTAC compounds. Wet experiments are conducted to validate the inhibitory effects of the compounds on the target as shown in Figure 1D.

## Data preprocessing for compounds and proteins

Data preprocessing, as shown in Figure 1A, involves the preprocessing of protein and small molecule compound data, including data filtering and molecular sequence segmentation.

The preprocessing of small molecule compounds begins with the selection of compounds according to the following rules: 1) Compilation of the source dataset based on the ZINC Clean Lead database; 2) Removal of molecules containing other electronegative atoms besides carbon, nitrogen, sulfur, oxygen, fluorine, bromine, and hydrogen; 3) Selection of drug-like compounds with a molecular weight between 200 and 600; 4) LogP (calculated using RDKit) ranging from -2 to 6. RDKit is used to convert small molecules into a unique representation in Canonical SMILES format. The obtained SMILES are parsed and split, and the positions containing side chains in the main chain are filled with the character 'R' to retain information about the side chains in the main chain. The positions where side chains are connected to the main chain are marked with "(", preserving the topological information of the side chains. The split SMILES are concatenated in the order of the main chain and side chains, forming a string-form SMILES. Subsequent steps are similar to protein splitting: the SMILES strings are segmented into *S-mol* using the VOLT algorithm, and a dictionary of *S-mol*s is constructed. *S-mol* with a frequency count below 5 are identified, and low-frequency masking is applied to them. During the encoding process, each SMILES is treated as a sentence, and each small *S-mol* is considered a word in this sentence. All small molecule SMILES are sequentially encoded, and after pretraining with Transformer, embeddings for each *S-mol* are obtained.

The preprocessing of protein involves concatenating the amino acid residue sequence according to the subunit order, thereby forming a complete amino acid residue sequence in that order. The residue sequence is segmented into *S-pro* using the VOLT algorithm, and a dictionary of *S-pro*s is constructed. *S-pro*s with a frequency count below 5 are identified as low-frequency *S-mol*s. During encoding, these *S-mol*s are all represented by the same *S-mol*, a method called low-frequency masking. During the encoding process, the encoding is performed according to the appearance order of *S-pro* in the dictionary. Here, each protein is treated as a sentence, and each *S-pro* is considered a word in this sentence. All proteins are sequentially encoded, and then processed through Transformer to obtain embeddings for each *S-pro*.

## Segmentation of molecular and proteins

The VOLT algorithm is employed in this work to seg proteins and molecules. VOLT is a vocabulary learning method based on Optimal Transport theory. The purpose of VOLT is to determine suitable vocabulary segmentation for specific tasks by leveraging optimal transport techniques, thereby enhancing the model's ability to

represent and understand vocabulary in NLP. The algorithmic process is outlined as follows:

Step 1, VOLT ranks all candidate *S-mol*s based on the pre-generated *S-mol* frequencies. For simplicity, VOLT typically employs *S-mol*s generated by BPE, such as BPE-100k, as candidates.

Step 2, All *S-mol*s with probabilities are used to initialize the optimal transport algorithm. At each time step, the vocabulary with maximum entropy can be obtained based on the transport matrix. Due to the relaxation strategy included in the optimal transport algorithm, situations of non-compliant transport may arise. Therefore, VOLT removes *S-mol*s with distribution frequencies less than 0.001.

Step 3, Exhaustively explore all time steps, selecting the vocabulary that satisfies the specified exponentiated search space as the final vocabulary.

Step 4, Use a greedy strategy similar to BPE to encode the text. Segment the sentence into character-level *S-mol*s. If the merged *S-mol* is present in the vocabulary, combine two consecutive *S-mol*s into one *S-mol* until no further merging is possible. *S-mol*s outside the vocabulary will be segmented into smaller *S-mol*s. If the *S-mol*s "CC(=O)NC" and "CC1=CN" are two adjacent *S-mol*s, and both are in the vocabulary, the sequence formed by concatenating them will be combined into a new *S-mol*, namely, "CC(=O)NC CC1=CN".

In order to ensure the acquisition of reasonable *S-mol* and to avoid including extremely small or excessively long *S-mol* in the *S-mol* library, further screening of SMILES *S-mol* was conducted based on the Chembridge approach. Supplementary Table S2 outlines the criteria used for filtering the *S-mol*s from the Chembridge Fragment Database during the construction of the *S-mol* library; *S-mol*s meeting these criteria were included in the library. Following the standards of the Chembridge Fragment Database, this experiment eliminated meaningless *S-mol*s and optimized the *S-mol* library to prevent the generation of invalid molecules caused by ineffective *S-mol*s.

## Screening for high SSI pairs

Extracting information from the protein-compound affinity library, the data is paired based on the embeddings obtained for *S-pro*s and *S-mol* in the previous step, generating pairs composed of *S-pro* and *S-mol*. In the design of PROTAC molecules, NLP technology aids in identifying key binding sites through the segmentation analysis of target proteins and *S-mol*s. Similar to word analysis in natural language, NLP processes these *S-mol* and generates efficient PROTAC molecules based on their interaction relationships. This approach reduces dependency on clearly defined binding pockets and expands the range of potential targets. The overall interaction between proteins and small molecule compounds is used to obtain local interactions, specifically the affinity between *S-pro* and *S-mol*. The affinity values from the protein-compound affinity library are used to train the FOTF-CPI prediction model.

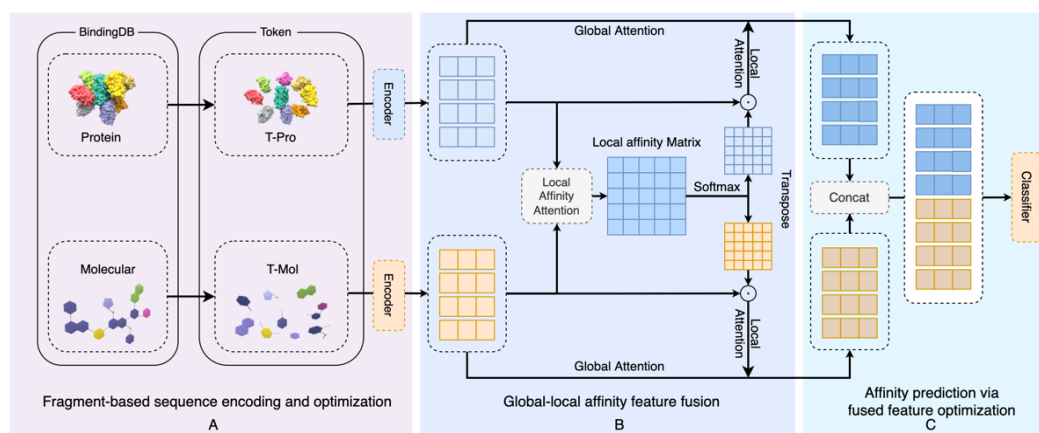

**Figure 2**. Mechanism of screening high SSI pairs

Step 1 fragment-based sequence encoding and optimization, the residual sequences of proteins are segmented into *S-pro*s using the VOLT algorithm, as shown in Figure 2A. The SMILES sequences of small molecule compounds are parsed and cut, and the cut SMILES sequences are concatenated in the order of main chain and side chain to form a string representation of SMILES. The protein sequence string is segmented into *S-mol* using the VOLT algorithm. The preprocessed sequences of small molecule compounds and protein sequences are randomly initialized as representations of small molecule compounds and proteins, respectively, based on *S-mol*. Similar to NLP methods, both the sequences of small molecule compounds and protein sequences are treated as complete sentences, with each *S-mol* considered as a word in the sentence. The representations of small molecule compounds and proteins, encoded in order based on fragments, are obtained separately after passing through an encoder. The entire network is continuously optimized based on the prediction results and real labels, using a combination of binary cross-entropy loss functions. The specific loss function is shown in the formula below:

$$loss = -\frac{1}{N}\sum_{i=1}^{N}(l_i \times \log(y_i) + (1 - l_i) \times \log(1 - y_i))$$

$y_i$ is the predicted result value, and $l_i$ is the true label value.

Step 2 global-local affinity feature fusion, the affinity relationship between fragments is obtained by multiplying the representations of small molecule compounds and proteins to generate a local affinity matrix as shown in Figure 2B. In order to avoid excessively high affinity scores in the local affinity matrix, normalization is applied, resulting in a new normalized local affinity matrix. The affinity matrix is then transformed through Softmax to obtain the affinity relationship matrix between each *S-mol* and different *S-pro*s. Simultaneously, by transposing the affinity matrix and applying Softmax, the affinity relationship matrix between each *S-pro* and different *S-mol*s is obtained. The product of the affinity relationship matrix and the representation of small molecule compounds yields the representation of small molecule compounds after local fragment correction. Similarly, the product of the affinity relationship matrix and the representation of proteins yields the representation of proteins. The representation of small molecule compounds after local fragment affinity attention correction is concatenated with the representation of small molecule compounds extracted under global attention correction in vector dimension, resulting in a mixed representation of small molecule compounds. Similarly, a mixed representation of proteins is obtained. The concatenated mixed representation of small molecule compounds and mixed representation of proteins, after passing through a fully connected layer and global adaptive pooling, respectively, yield the representations of small molecule compounds and proteins after fusion of global and local features.

Step 3 affinity prediction via fused feature optimization, concatenate the fused representations of small molecule compounds and proteins in the fragment dimension to obtain affinity features as shown in Figure 2C. The obtained affinity features are then sequentially passed through a global adaptive pooling layer and an activation function layer to obtain a pair of predictions for small molecule compounds and proteins. Based on the predicted results and real labels, continuously optimize the entire network using a binary cross-entropy loss function.

## Molecular generation based on structural and molecular property constraints

In this research, a Conditional Transformer is employed the physicochemical properties of the generative model. Within the foundational model of the Conditional Transformer, constraints based on two attributes are introduced to guide the molecular generation process. This approach serves to constrain crucial properties during the molecular generation process, enhancing efficiency. Simultaneously, reducing the number of introduced attributes helps avoid an excessive increase in model size due to an overabundance of parameters, thereby reducing both training and operational costs of the model.

The molecular generation model is based on the C-Transformer model. Similar to the *S-pro* pretraining model, it involves molecular segmentation for encoding and training. The training data is derived from Chembl and ZINC, consisting of 250k selected drug-like small molecules. In the encoding and generation embedding process, the Transformer is used to extract long-range interactions between compound atoms. This C-Transformer generation model is trained using small molecule data with labeled properties (LogP and LogSW). Throughout the model training iterations, the labeled properties serve as one of the conditional encodings for the C-Transformer, while the molecular SMILES are used for structural encoding training. Simultaneously, C-Transformer Embedding is employed to encode a randomly initialized molecule and any *S-mol*s, calculating the distance $D$ between the initial molecule and the *S-mol*s. The similarity $S$ between the molecular skeleton of the random initial molecule and the *S-mol*s skeleton is computed. Reinforcement learning is applied, using the product of $D$ and $S$ as a reward reference. Rewards are given for higher similarity and closer distance, while penalties are imposed inversely, thereby constraining the generated molecular structure. The loss value of the model is determined by the following parameters:

$$loss = -\frac{1}{n}\sum_x [a \ln D * S + (1-a)\ln(1 - D*S)]$$

$D$ represents the Euclidean distance between the initial molecule and the *S-mol*, and $S$ represents the similarity between the molecular skeleton and the *S-mol* molecular skeleton. Parameter $a$ is used to allocate the impact of C-Transformer and reinforcement learning on the model gradients, while $n$ represents the batch size.

Tanimoto similarity is a method used to measure the similarity between two sets, commonly applied in cheminformatics to compare the similarity of molecular fingerprints[61]. The formula for calculating the Tanimoto similarity used in this article is as follows:

$$\text{TanimotoSimilarity} = \frac{A \cap B}{A \cup B}$$

Where $A \cap B$ denotes the number of elements in the intersection of sets $A$ and $B$.

The specific calculation steps are as follows:

Step 1 molecular fingerprints representation, represent molecules as binary vectors (a sequence of 0s and 1s), where each bit indicates whether a particular feature exists in the molecule.

Step 2 intersection calculation, perform a bitwise "AND" operation on the binary vectors of the two molecules to determine the number of bits that are 1 in both, which represents the intersection.

Step 3 union calculation, perform a bitwise "OR" operation on the binary vectors of the two molecules to determine the number of bits that are at least 1, which represents the union.

Step 4 similarity calculation, divide the number of elements in the intersection by the number of elements in the union to obtain the Tanimoto similarity.

The range of Tanimoto similarity is from 0 to 1; a value closer to 1 indicates that the two molecular fingerprints are more similar, while a value closer to 0 indicates that they are less similar.

Simultaneously, molecular properties such as lipophilicity and water solubility will be incorporated into the constraints of the molecular generation model to ensure that the generated PROTAC molecules exhibit greater diversity in these properties. During the generation process, in the step where the generated molecules undergo *S-mol* structure constraints, a fixed molecular *S-mol* library is used, consisting of screened *S-mol* with high affinity for the target protein. Additionally, the combined *S-mol*s of these *S-mol*s are also added to the *S-mol* library. The generated PROTAC molecules undergo a synthetic feasibility check using RDKit. If molecules with unsatisfactory synthetic feasibility are generated, adjustments are made strictly according to the SMILES syntax to improve the compliance rate of the molecules.

**Performance evaluation of generative models**

**Generative performance**

The general statistical metrics aim to assess the model's ability to generate new molecular structures. The validity metric is used to evaluate the model's capability to generate chemically valid molecules. Valid molecules are those that correspond to some real-world molecules, at least theoretically. The validity is calculated as the ratio of valid molecules $N_{valid}$ to all generated molecules $N_{generated}$. Molecules with incorrect SMILES syntax or invalid bond values will be penalized [62].

$$Vaildity = \frac{N_{valid}}{N_{generated}} * 100\%$$

The novelty metric is used to evaluate the model's ability to generate new molecules. A model should have good coverage of chemical space to generate molecules that are similar but not identical to those in the training set[63]. The novelty is calculated as the ratio of new molecules to all valid molecules, where $N_{existed}$ is the number of molecules generated that are duplicates with the training set, and $N_{valid}$ is the total number of generated molecules. Molecules already present in the training set will be penalized [62].

$$Novelty = (1 - \frac{N_{existed}}{N_{valid}}) * 100\%$$

The uniqueness metric is used to assess the model's ability to generate distinct molecules. A model should explore the entire chemical space to generate a large number of unique molecules, rather than lazily generating duplicates to reduce loss. The uniqueness is calculated as the ratio of unique molecules $N_{unique}$ to all valid molecules $N_{valid}$. Duplicated generated molecules will be penalized[62].

$$Uniqueness = \frac{N_{unique}}{N_{valid}} * 100\%$$

**Property distribution**

The distance between the chemical property distributions of generated molecules and molecules in the training set is considered useful for assessing how well the model has learned from the training set. Two metrics have been proposed for this purpose[62].

FCD was introduced by Preuer et al. [64]. They introduced and trained a neural network called ChemNet to predict biological activity. The latent vectors from the penultimate layer of ChemNet are extracted, and the averages and covariances of these activations for both the reference set and the generated set of molecules are computed. The Fréchet distance between the obtained pairs of values is then calculated as FCD.

**KL divergence**

KL divergence is a crucial metric for measuring how one probability distribution approximates another. In this benchmark test, physical-chemical properties, including BertzCT, MolLogP, MolWT, Topological Polar Surface Area (TPSA), NumHAceptors, NumHDonors, NumRotatableBonds, NumAliphaticRings, and NumAromaticRings, are computed for generated molecules and the training set using RDKit tools. The distributions of these descriptors are then calculated. For a total of N descriptors, the Kullback-Leibler divergence $D_{KL\ i}$ is computed for each descriptor i and aggregated to obtain the final score $S$.

$$S = \sum_{i=1}^{N} D_{KL\ i}$$

**Attribute filtering**

Establishing the small molecule screening model MDAM involves selectively constraining molecular features that meet the requirements of drug molecules, aiming to filter out potential PROTAC candidate molecules that meet the criteria. Property screening includes two aspects: affinity repetition screening, which involves experimentally calculating the affinity between generated molecules and target molecules, and other property screening, including lipophilicity, water solubility, etc., to satisfy specific drug release properties. MDAM is a multi-dimensional attribute fusion model constructed by sampling 1D, 2D, and 3D information of molecules. Particularly, the sampling of the 3D structure of molecules utilizes SphereNet and attention mechanisms to generate a 3D structure feature network, obtaining the feature vector of the molecule's 3D structure through pre-training.

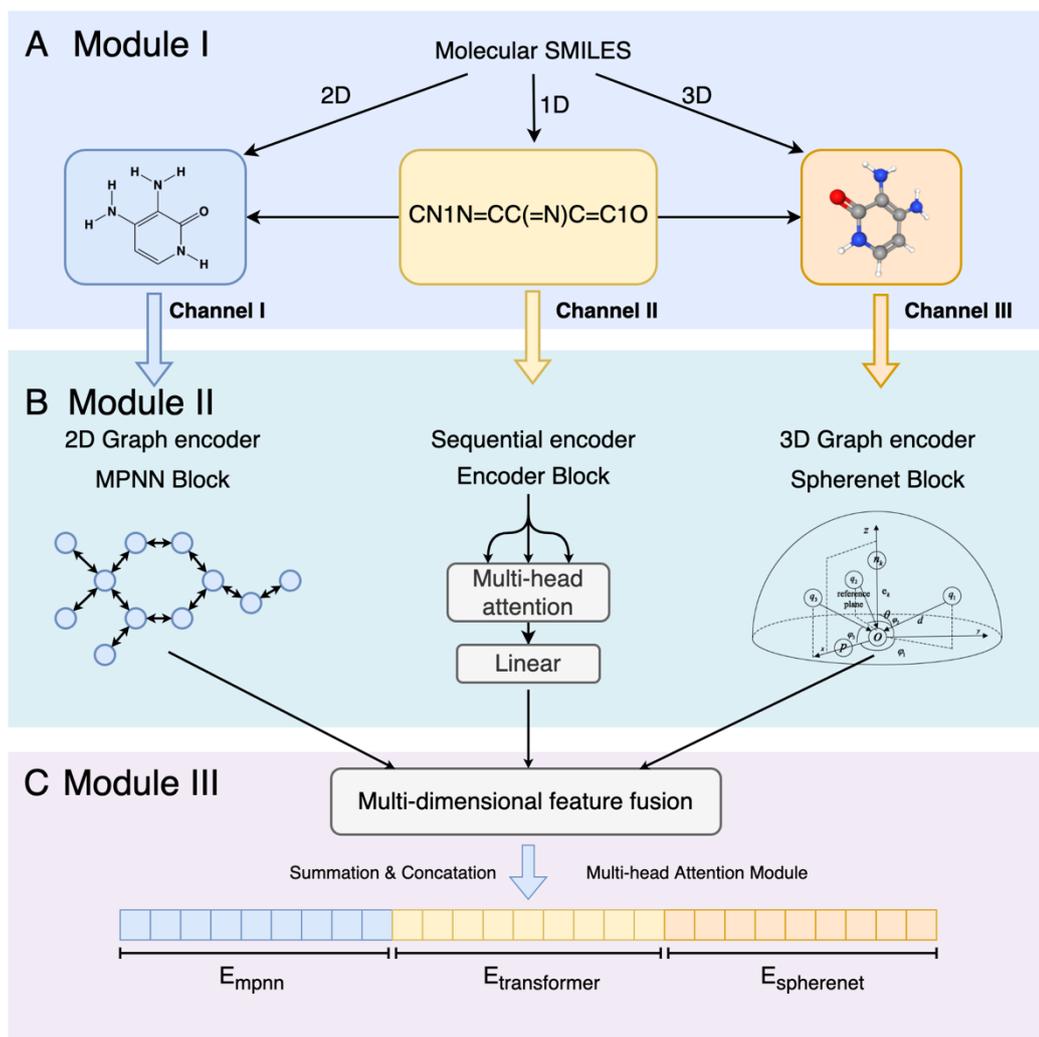

**Figure 3.** The process of the MDAM based on attention mechanism

To screen for PROTAC molecular properties, a PROTAC molecular property prediction model based on MDAM model was developed. The model consists of three main modules:

Module I: Multi-Dimensional Feature Encoding Module as shown in Figure 3A.

Module II: Generation of 1D, 2D, and 3D feature vectors for molecules as shown in Figure 3B.

Module III: Multi-Dimensional Feature Fusion Module as shown in Figure 3C.

The attention-based multi-dimensional feature encoder takes the SMILES-encoded model of a molecule as input and is divided into three channels.

Channel I: Processing the SMILES string, MDAM utilizes the FCS algorithm to directly encode the SMILES, generating sequence data as input for the sequence encoder. The sequence encoder, employing a Transformer, directly encodes the sequence data to obtain the feature vector for the molecular 1D sequence.

Channel II: RDKit transforms the SMILES string into a molecular graph, serving as input for the graph encoder. Through an attention layer, the graph encoder calculates attention scores for adjacent nodes as weighting coefficients while aggregating neighboring message vectors. This process ultimately yields the molecular 2D atomic graph feature vector.

Channel III: Handling structural information such as 3D coordinates of the molecule, MDAM utilizes SphereNet and a multi-head attention mechanism to generate the feature vector for the molecular 3D structure. After obtaining feature vectors from the three channels - 1D sequence feature vector, 2D atomic graph feature vector, and 3D structure feature vector - the Feature Fusion Module is used to merge these vectors and predict downstream molecular properties.

$$\text{MultiHead}(Q, K, V) = \text{Concat}(\text{head}_1, \text{head}_2, \ldots, \text{head}_m)W^o$$

$$\text{Head}_i = \text{Attention}(Q \times W_i^Q, K \times W_i^K, V \times W_i^V) = \text{softmax}(\frac{Q_i \times K_i^T}{\sqrt{d_k}})V_i$$

$$X = \text{concat}(E_{\text{transformer}}^{(M_i)}, E_{\text{mpnn}}^{(M_i)}, E_{\text{spherenet}}^{(M_i)}) = (E_{\text{transformer}}^{(M_i)}, E_{\text{mpnn}}^{(M_i)}, E_{\text{spherenet}}^{(M_i)})$$

Where $W_i^Q$, $W_i^K$ and $W_i^V$ are the weight matrices for each attention head, and $Q_i$, $K_i$ and $V_i$ are the query, key, and value vector matrices derived from linear transformations, respectively. The attention operation represents the calculation process of a single attention head, where attention weights are computed based on the query, key, and value, followed by a weighted sum.

$E_{\text{spherenet}}^{(M_i)}$ denotes the 3D molecular vector obtained from SphereNet, $E_{\text{mpnn}}^{(M_i)}$ represents the 2D molecular vector obtained from MPNN, and $E_{\text{transformer}}^{(M_i)}$ is the 1D molecular vector obtained from the transformer.

In downstream tasks, there are both regression tasks and classification tasks. In classification tasks, the output of the decoder is a result, while in regression tasks, the output is a predicted value. MDAM is optimized using both cross-entropy loss function and MAE loss function. The loss function in this model is represented by the formula:

$$loss_{CE} = -\sum NlogN - \lambda(1-N)\log(1-N)$$

$$loss_{MSE} = \frac{1}{M}\sum_{i=1}^{M}(y_i' - y_i)^2$$

Where $y_i$ is the class label of the sample, and $y_i'$ is the predicted value of the sample.

In the processes of chemical synthesis and drug development, the SAS and the QED are two commonly used screening tools for selecting molecules that are easier to synthesize and possess drug-like properties. The threshold setting for these two indicators during the screening process is crucial, as they directly impact the quality of the screened molecular library and the success rate of subsequent experimental studies. In this article, SAS and QED are used as final selection criteria for further filtering the generated PROTAC molecules.

SAS is a quantitative metric used to assess the synthetic difficulty of a molecular structure, with scores typically ranging from 1 to 10, where lower scores indicate that a molecule is easier to synthesize[65]. The calculation of SAS is based on various structural features of the molecule, including molecular complexity, the number and types of chemical groups, the size and number of rings, and stereochemical features. During the molecular screening process, researchers often set a SAS threshold based on specific project requirements and the synthetic capabilities of their laboratory. For example, in a project with limited synthetic resources, they might choose to retain molecules with an SAS score below 3, indicating that these molecules have high synthetic accessibility and are easier to synthesize under laboratory conditions. Conversely, in situations where synthetic resources are abundant or the target molecules have higher potential efficacy, the threshold may be relaxed to around 5. Generally, molecules with an SAS score greater than 6 are considered to have high synthetic difficulty, potentially requiring more complex synthetic routes, and are thus usually excluded from priority screening.

First, SAS is based on the decomposition of the molecular structure, breaking the entire molecule down into several *S-mol*. The complexity of each *S-mol* is determined by a series of chemical features, including atom types, bond types, the number and size of ring systems, the number and types of functional groups, as well as stereochemical information (such as the presence of chiral centers). Specifically, for each *S-mol*, the complexity score $C_i$ can be represented as:

$$C_i = f(Atom\ Type, Bond\ Type, Ring\ Structure, Stereochemistry)$$

Here, the function $f$ represents a comprehensive evaluation of these features. After completing the complexity assessment of the *S-mol*s, the calculation of SAS continues by aggregating the complexity scores of all *S-mol*s, while also considering the overall structural characteristics of the molecule, such as the connections between *S-mol*s $L_j$ and the symmetry $S$ of the molecule. These factors influence the overall synthetic difficulty, so when calculating the total molecular complexity score $C_{total}$, the following formula is typically used:

$$C_{total} = \sum_{i=1}^{n} w_i \times C_i + \sum_{j=1}^{m} g(L_j) + h(S)$$

Here, $w_i$ is the weight of each *S-mol*, $L_j$ is a function that accounts for the contribution of the connection between *S-mol*s to the overall complexity, and $h(S)$ is a correction function that considers the influence of molecular symmetry on the complexity. Since the contributions of different *S-mol*s may be uneven, the weights $w_i$ allow for amplifying or reducing the complexity of certain *S-mol*s.

Finally, the overall complexity score $C_{total}$ is normalized to a range of 1 to 10 to ensure consistency and comparability of SAS scores across different molecules. The normalization typically uses a linear transformation formula:

$$SAS = 1 + 9 \times \frac{C_{total} - C_{min}}{C_{max} - C_{min}}$$

Here, $C_{min}$ and $C_{max}$ represent the lowest and highest complexity scores among known molecules, respectively. This normalization method ensures that SAS maintains a relatively consistent evaluation scale across different molecules, where a score of 1 indicates that the molecule is very easy to synthesize, and a score of 10 indicates that it is very difficult to synthesize.

QED is a quantitative metric used to evaluate the drug-like properties of a compound, calculated based on a combination of several key molecular characteristics. These characteristics include molecular weight, LogP, HBD, HBA, PSA, NRB and AR. The calculation of QED first involves determining the score for each characteristic using the following formula:

$$d_i = e^{-\left(\frac{X_i - \mu_i}{\sigma_i}\right)^2}$$

Here, $X_i$ represents the actual value of characteristic $i$, while $\mu_i$ and $\sigma_i$ are the mean and standard deviation of characteristic $i$, based on statistical data from known drug molecules. Each characteristic score $d_i$ is normalized to a range from 0 to 1, where a value of 1 indicates that the characteristic fully aligns with drug-like properties, and 0 indicates a complete lack of alignment.

Then, the weighted geometric mean of all normalized characteristic scores $d_i$ is used to calculate the QED, as given by the following formula:

$$QED = \prod_{i=1}^{n} d_i^{w_i}$$

Here, $w_i$ represents the weight of characteristic $i$, reflecting the relative importance of that characteristic in evaluating drug-likeness. In this way, QED integrates multiple key characteristics of a molecule to generate a single score ranging from 0 to 1, with a value closer to 1 indicating a stronger drug-like nature.

During the screening process, the QED threshold is typically determined based on the drug property requirements of the target application. A QED threshold of 0.5 is often set to retain as many potential candidate molecules as possible for further screening. However, in more stringent screening stages, the QED threshold is usually set at 0.7 or higher, ensuring that the retained molecules have a higher potential for drug-likeness, showing favorable pharmacokinetic ADME and toxicity profiles. In general, molecules with a QED score below 0.3 are considered unlikely to possess ideal drug-like properties and are therefore often excluded from further studies[66].

By combining SAS and QED for molecular screening, both the synthetic feasibility and drug-likeness of the molecules can be considered, effectively filtering out compounds that are difficult to synthesize or lack desirable drug-like properties at an early stage. For PROTAC molecules, based on synthesis difficulty and cost, molecules with an SAS score below 3 are retained. In the case of early-stage candidate compounds, QED does not need to be a strict criterion, so a threshold of 0.5 is set for QED screening. This screening process, which integrates both metrics, helps to conserve synthesis resources while also enhancing the drug-like properties of the molecules.

**Molecular docking**

**Protein structure and ligand structure acquisition**

Uniprot is a protein database that includes protein sequences, functional information, and research paper indexes, integrating resources from three major databases: EBI, SIB, and PIR. Using "WNT3A" as the keyword, protein structures were retrieved from Uniprot, selecting the EM structure of WNT3A_HUMAN (Uniprot ID: P56704) with a resolution of 2.20 Å (PDB ID: 7DRT). Subsequently, its ligand SMILES were converted into PDB format for preservation.

**Protein and ligand preparation**

The downloaded protein PDB file was imported into Schrödinger Maestro 13.1 software, and the remaining chains of the protein as well as water molecules and impurities were removed. Chain A was retained, and the Protein Preparation module was used to check for any missing atoms or residues in the protein crystal structure and correct them accordingly. Hydrogen atoms were added to the protein structure, and partial charges were assigned to each atom. Compounds were subjected to energy minimization using the LigPrep module, with the OPLS4 force field selected, and the processed compounds were saved for later use.

### Grid generation for docking

The "SiteMap" module was first used to predict potential pockets on the protein, and after selecting the optimal binding pocket, the Receptor Grid Generation module was used to generate the docking box.

### Molecular docking

The prepared ligand library was docked to the protein's active site using the precision XP mode in the Ligand Docking module of the Schrödinger suite. The Glide-score value, representing the docking score of the compound to the target, was obtained. Glide-score is an empirical scoring function designed to maximize the separation between compounds with strong binding affinity and those with little or no binding capability. It consists of terms describing the physical properties of the binding process, including hydrophobic-hydrophobic interactions, hydrogen bonding interactions, and contributions of protein-ligand Coulomb-van der Waals energy. A higher Glide-score indicates a more stable binding of the ligand to the receptor. The binding activity of compounds to the target was evaluated based on the Glide-score value. Three ligands were docked to the protein individually.

### Analysis of protein-ligand interactions

The binding interface of the protein-ligand complex was systematically analyzed using PLIP, and interaction-related details were supplemented using pyMOL 2.5 software.

Hydrogen bonds are considered the most important among all directional covalent interactions. The upper limit for their distance at was configured as 4.1 Å. Hydrogen bonds are formed between a donor group (such as DH) providing a hydrogen atom in the form of a positive end and an acceptor group with high electron density. With each additional hydrogen bond, the binding affinity of the ligand increases by an order of magnitude. The typical range of hydrogen bond energies is 10 to 40 kJ/mol.

**Molecular dynamics simulations**

The MD simulations were conducted to study the interactions between three PROTAC candidate compounds within the Wnt3a and CRBN complex system. The simulations were run using GROMACS on a server equipped with an Intel E5 2686 v4 processor running at 2.30 GHz and an NVIDIA 3090 Ti GPU with 24 GB VRAM, using the Ubuntu 20.04 operating system. Each MD simulation was conducted over a total duration of 500 ns.

In this study, the initial complex structure of Wnt3a, CRBN, and each of the three PROTAC candidate molecules was prepared and parameterized for MD simulation. The system was solvated in a cubic box with TIP3P water molecules, and appropriate ions were added to neutralize the system. Energy minimization was performed to relax the system, followed by equilibration phases under the NVT and NPT ensembles to stabilize temperature and pressure. Production MD runs were conducted under periodic boundary conditions with a time step of 2 fs, and simulations were carried out with a 12 Å cutoff for nonbonded interactions and the PME method for long-range electrostatics. Complex visualization and rendering were performed using PyMOL version 3.0.3.

During the simulation analysis, several key metrics were evaluated:

RMSD values for both the protein and ligand were calculated to assess the stability of the protein-ligand complex over time, providing insights into the structural integrity and dynamic behavior of each candidate compound within the Wnt3a-CRBN complex.

The binding free energy $\Delta G$ of the complexes was calculated using the MM-GBSA method, allowing the identification of key residues contributing to the binding affinity for each candidate compound.

The Rg values were monitored to evaluate the overall compactness and structural stability of the protein-ligand complexes throughout the simulation.

The RMSF analysis was performed to determine the flexibility of specific residues within the protein complex, highlighting regions with high mobility.

The SASA values were calculated over time to observe changes in the exposure of the protein-ligand interface to the solvent, reflecting potential changes in the complex's stability and interactions.

**Compound synthesis process**

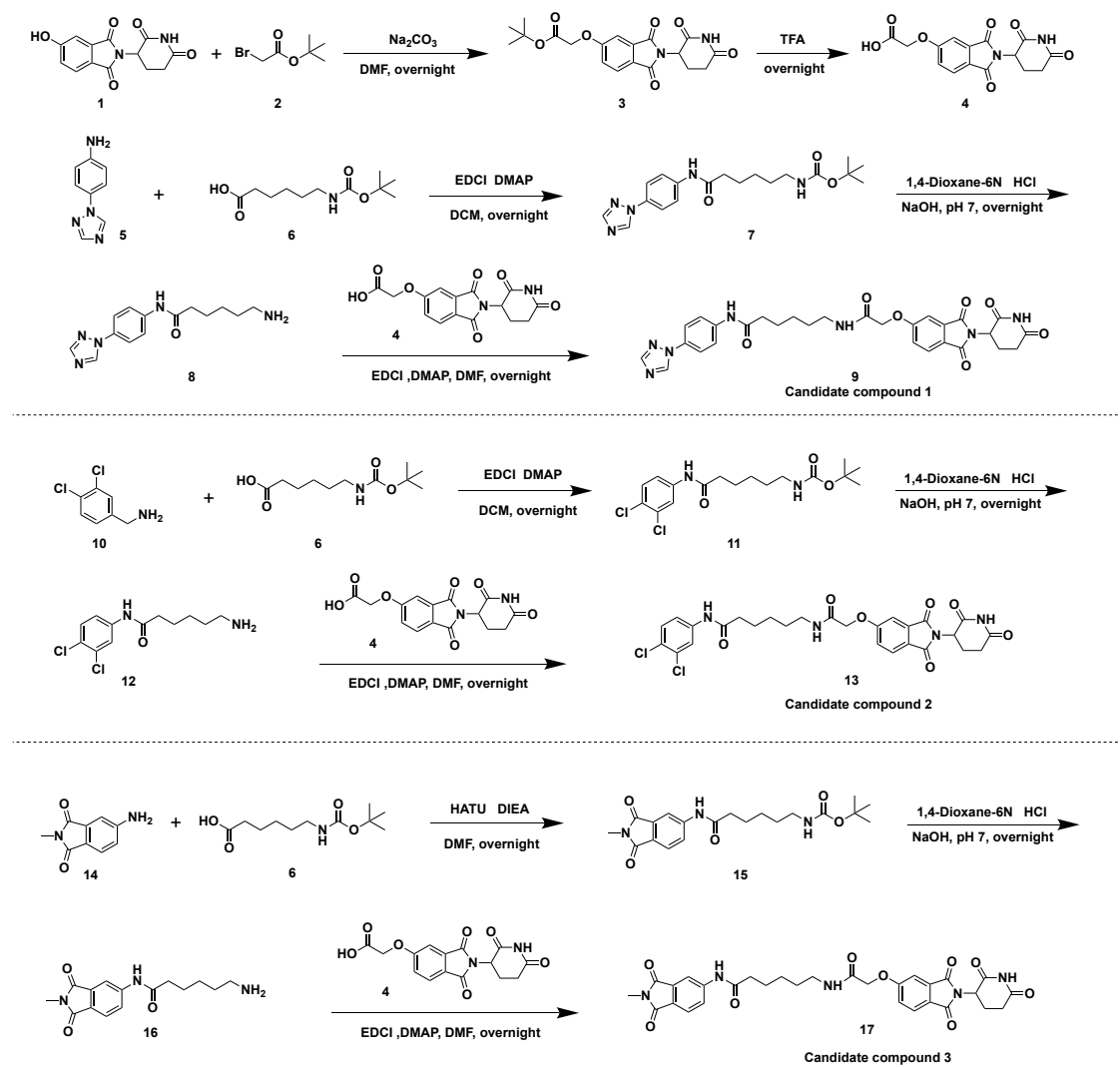

Figure 4: Synthetic Pathway for Candidate Compound 1-3.

Figure 4 illustrates the synthetic pathway of Compound 1-3, with detailed steps as follows:

A mixture solution of 1 (2.5 g, 9 mmol) and Na2CO3 (1.9 g, 18 mmol) in DMF was stirred for 30 minutes, and then 2 (2.1 g, 11 mmol) was added and stirred at room temperature overnight to obtain 3. 3 was further reacted with DCM and TFA to give 4 by removing the Boc-protecting group. To obtain 7, 5 (1.6 g, 10 mmol), 6 (2.3 g, 12 mmol) and DMAP (0.12 g, 1 mmol) were added with EDCI (2.3 g, 12 mmol) to DCM, and then the mixture was stirred at room temperature overnight. To a solution of 7 (1.5 g, 4 mmol) in 1,4-Dioxane at room temperature, 6N HCl was added and stirred overnight. After completion of the reaction, the mixture was quenched to pH 7 with NaOH to produce the product 8. 8 (300 mg, 1.1 mmol), 4 (365 mg, 1.1 mmol) and DMAP (16 mg, 0.132 mmol) were added with EDCI (253 mg, 1.32 mmol) to DCM, and then the mixture was stirred at room temperature overnight. The resulting mixture was extracted with EtOAc (3 × 10 mL), dried over $Na_2SO_4$, filtered, and then concentrated under a reduced pressure. The residue was purified by flash silica chromatography (petroleum ether/ethyl acetate = 2:1 v/v) to obtain candidate compound 1.

Employing the same reaction conditions as above, the residue was purified by flash silica chromatography (dichloromethane/methyl alcohol = 15:1 v/v) to obtain candidate compound 2.

14 (1.76 g, 10 mmol), 6 (2.78 g, 12 mmol) and HATU (5.7 g, 15 mmol) were added with DIEA (2.58 g, 20 mmol) to DMF in an ice-water bath, and then the mixture was stirred at room temperature for 4h to obtain 15. And then employing the same reaction conditions as above, the residue was purified by flash silica chromatography (petroleum ether/ethyl acetate = 4:1 v/v) to obtain candidate compound 3.

**In vitro validation**

**Cell Culture and Reagents:** HepG2 human hepatocellular carcinoma cells were cultured in DMEM high-glucose medium (Shanghai Titan Technology, China) containing 10% fetal bovine serum (FBS, Lonsera), 1% penicillin and streptomycin

(Shanghai Beyotime Biotechnology, China). Cells were incubated at 37°C with 5% $CO_2$ in a Thermo Fisher cell incubator (Thermo Fisher, USA). Compounds were dissolved in DMSO (Sigma, USA) to prepare an 80 mM stock solution.

**Cell Treatment**: Once HepG2 cell density reached 85–90% confluence, the medium was removed, and cells were washed with 2 mL of phosphate-buffered saline (PBS, BBI). Cells were digested with 1 mL of 0.25% trypsin (ScienCell) at room temperature for 1–2 minutes. After observing sufficient detachment, cells were collected, centrifuged at 1000 rpm for 5 minutes, and resuspended in 1 mL of complete medium. Cells were seeded in 12-well plates at an appropriate density and incubated for 24 hours at 37°C with 5% $CO_2$. After incubation, the medium was removed, and cells were treated with DMEM containing various concentrations of the compounds (80 μM, 40 μM, 20 μM, 10 μM, 5 μM, 2.5 μM, 1.25 μM, 0.625 μM, 0.3125 μM, and 0.15625 μM). Following 24 hours of treatment, the medium was removed, and cells were washed twice with 1 mL of sterile PBS for downstream experiments.

**Protein Extraction and Quantification**: For protein extraction, cells were lysed with RIPA buffer (Shanghai Beyotime Biotechnology, China) containing protease inhibitor PMSF (Shanghai Beyotime Biotechnology, China) and incubated on ice for 5–10 minutes. Cells were scraped and collected into 1.5 mL microcentrifuge tubes. Lysates were centrifuged at 12000 rpm for 15 minutes at 4°C, and the supernatant was collected. A portion of the protein extract was used for protein concentration determination using the BCA Protein Assay Kit (Shanghai Beyotime Biotechnology, China). The remaining protein was mixed with 5× loading buffer and boiled at 98°C for 10 minutes.

**SDS-PAGE and Western Blot Analysis**: Proteins were separated using 10% SDS-PAGE, with 5 mL of resolving gel and 2 mL of stacking gel prepared per plate. Electrophoresis was conducted at 80 V for 40 minutes, followed by 120 V for approximately 1 hour until complete protein separation. Following electrophoresis, proteins were transferred to a PVDF membrane (Millipore, USA) using a 260 mA current for 70 minutes. The membrane was blocked with a quick-blocking buffer for 10–20 minutes at room temperature, then incubated overnight at 4°C with primary antibodies: Wnt3a (1:1000) and GAPDH (1:10000) (both from Saiye Biotech, China). The membrane was washed with TBST and incubated with HRP-conjugated secondary antibodies (HRP-conjugated Goat Anti-Mouse IgG and HRP-conjugated Goat Anti-Rabbit IgG, Saiye Biotech, China) for 2 hours at room temperature. After further washes, the membrane was incubated with ECL reagent (Tanon, China) and visualized using a Tanon imaging system.

## Result and discussion

### Preprocessing of molecular and protein data

This experiment utilized a dataset sourced from ZINC for molecular *S-mol* splitting, which resulted in a total of 4,577,207 SMILES *S-mol*s. To ensure the acquisition of reasonable *S-mol* segments and to avoid including extremely small or excessively long segments in the *S-mol* library, further screening of SMILES *S-mol*s was conducted based on the Chembridge approach, resulting in 870,834 *S-mol*s after the screening process.

Protein splitting data were obtained from the BindingDB dataset. The VOLT method was employed to split the protein data. After the protein splitting process, the obtained *S-pro*s underwent low-frequency masking treatment to create the *S-pro* collection.

The primary purpose of applying low-frequency masking to a molecular *S-mol* library was to improve the model's generalization ability and robustness. Low-frequency molecular *S-mol*s appeared infrequently in the training data, making it difficult for the model to learn stable and useful features from these rare *S-mol*s. By masking these low-frequency *S-mol*s, the model could focus more on high-frequency *S-mol*s and learn more generalized features, leading to better performance when handling new data. Additionally, low-frequency *S-mol*s often contained more noise and uncertainty, and if the model overly relied on these *S-mol*s, it could lead to overfitting. Masking low-frequency *S-mol*s reduced the model's dependence on this noisy data, thereby improving the model's robustness and stability[30].

Low-frequency *S-mol*s were often numerous but contributed little to the overall dataset. By masking these low-frequency *S-mol*s, the training process was simplified, reducing computational costs and improving training efficiency. The infrequent appearance of molecular *S-mol*s led to data sparsity, which negatively impacted the model's learning effectiveness. Masking low-frequency *S-mol*s helped alleviate the data sparsity issue, enabling the model to more effectively learn and represent molecular *S-mol*s. In practical applications, the model might have encountered many unseen molecular *S-mol*s. By masking low-frequency *S-mol*s during training, the model adapted better to these new *S-mol*s, improving its generalization performance and robustness. At the same time, the low-frequency masking strategy indirectly increased the probability of less frequent *S-mol*s, promoting molecular diversity while injecting more randomness into the sampling process.

**Filtering high SSI**

Table 1: Performance of FOTF-CPI

| Model | AUC$_{(SD)}$ | PRC$_{(SD)}$ | Sensitivity$_{(SD)}$ | Specificity$_{(SD)}$ | F1$_{(SD)}$ | Cost$_{(h)}$ |
|---|---|---|---|---|---|---|
| DeepDTA[67] | 0.901$_{(0.007)}$ | 0.810$_{(0.006)}$ | 0.780$_{(0.024)}$ | 0.905$_{(0.017)}$ | 0.757$_{(0.014)}$ | 12.65 |
| TransformerCPI[68] | 0.910$_{(0.007)}$ | 0.788$_{(0.011)}$ | 0.736$_{(0.014)}$ | 0.890$_{(0.012)}$ | 0.731$_{(0.005)}$ | 5.44 |
| Moltrans[69] | 0.903$_{(0.002)}$ | 0.806$_{(0.007)}$ | 0.762$_{(0.013)}$ | 0.908$_{(0.007)}$ | 0.752$_{(0.004)}$ | 1.20 |
| ML-DTI[70] | 0.902$_{(0.007)}$ | 0.785$_{(0.011)}$ | 0.753$_{(0.007)}$ | 0.851$_{(0.005)}$ | 0.763$_{(0.003)}$ | 2.32 |
| IIFDTI[71] | 0.917$_{(0.003)}$ | 0.793$_{(0.004)}$ | 0.817$_{(0.011)}$ | 0.883$_{(0.013)}$ | 0.745$_{(0.003)}$ | 18.9 |
| FOTF-CPI | *0.929*$_{(0.003)}$ | *0.834*$_{(0.002)}$ | *0.822*$_{(0.006)}$ | *0.924*$_{(0.012)}$ | *0.789*$_{(0.005)}$ | 0.98 |

Table 1 presented a comparison study of the BindingDB dataset based on AUC, PRC, sensitivity, specificity, and F1. Notably, the FOTF-CPI proposed in this research obtained the highest scores on each of the metrics. A comprehensive comparative study of the Davis and Biosnap datasets was provided in the Moltrans study[69]. This reference offered a detailed evaluation of the performance of various models, including FOTF-CPI and ML-DTI, across several critical metrics such as accuracy, precision, recall, and F1-score. For the Davis dataset, FOTF-CPI demonstrated superior performance on most metrics, with the exception of specificity, where ML-DTI excelled. Similarly, in the Biosnap dataset, FOTF-CPI achieved the highest scores in most evaluation criteria, except sensitivity, where ML-DTI performed better. This comparative analysis highlighted the strengths and trade-offs of these models in predicting drug-target interactions.

Although FOTF-CPI had slightly lower specificity and sensitivity on the Davis and Biosnap datasets alone, it remained the best model among several models when both specificity and sensitivity metrics were considered together. The DeepDTA[67], TransformerCPI[68], Moltrans[69], ML-DTI[70], and IIFDTI[71] models focused on interactions between proteins and compounds as entire sequences, while the interactions between protein and compound *S-mol*s were not examined as essential components. The FOTF-CPI model prioritized the collection of sequence characteristics of molecular *S-mol*s and then mined them for information links. As a result, more feature information was acquired, and the performance of CPI was enhanced.

**Generated molecules**

A structure-constrained molecular generation model for comparison with other methods was trained and evaluated.

Table 2: Performance of DCT model

| Generation Model | Validity | Uniqueness | Novelty | FCD score | KL divergence |
|---|---|---|---|---|---|
| CharRNN[72] | 0.975 | 0.999 | 0.842 | 0.913 | 0.991 |
| AAE[73] | 0.937 | 0.997 | 0.793 | 0.555 | / |
| VAE[74] | 0.937 | 0.998 | 0.695 | 0.099 | 0.567 |
| JT-VAE[37] | 1.000 | 0.999 | 0.914 | 0.395 | 0.822 |
| Organ[75] | 0.961 | 0.923 | 0.872 | 0.000 | 0.267 |
| MolGPT[76] | 0.994 | 1.0 | 0.797 | 0.067 | 0.507 |
| DCT | 0.998 | 0.999 | 0.997 | 0.083 | 0.921 |

Table 2 compared the performance of the DCT generation model with other similar models. The experimental results were obtained by generating new molecules based on the MOSES dataset to calculate performance parameters[77]. From Table 2, it was observed that the FCD value of our model was 0.083, indicating a relatively low level, which suggested that the model successfully captured the statistical characteristics of the dataset. Additionally, the high KL divergence value of 0.921 for DCT, which was close to those of other advanced models, suggested that DCT could correctly generate molecules with features that aligned with the original dataset distribution.

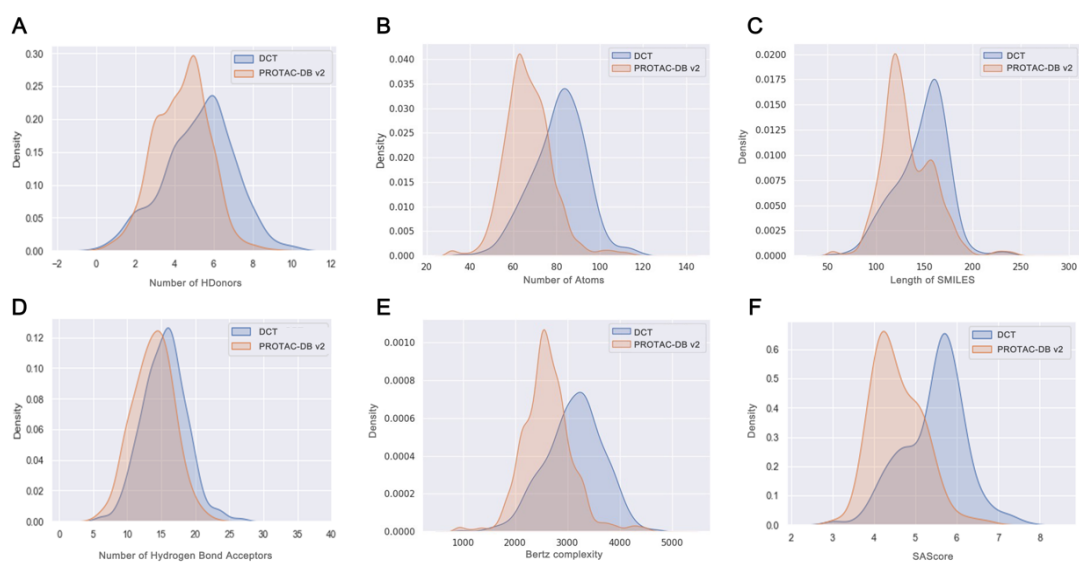

**Figure 5.** The attributes distribution of generated model outputs and the training set.

As shown in Figure 5, a generative model using the PROTAC-DB dataset was trained. The trained generative model was then utilized to generate 10,000 molecules from scratch. The various properties of these 10,000 generated molecules were statistically analyzed, and corresponding distribution plots were generated. In each distribution plot, the property distributions calculated from the original PROTAC-DB 2.0 dataset were included. From the plots, it was evident that the generative model accurately defined the chemical space of the input training set. In the distributions of multiple properties, including Number of HDonors, Number of Atoms, Length of SMIELS, Number of Hydrogen Bond Acceptors, Bertz complexity, and SAScores, the distribution range of the generated molecules closely matches that of the molecules in the original dataset.

The distribution of hydrogen bond donors in the molecules generated by DCT was similar to that of the molecules in the training set, with a peak at 6, which was 1 more than that in PROTAC-DB v2, as seen in Figure 5A. Considering that strict constraints on the water solubility of the generated molecules were applied in the generative model, molecules with more hydrogen bond donors generally exhibited higher water solubility as shown in Figure 5B. Therefore, the constraint on water solubility led to higher values of hydrogen bond donors in the generated molecules, causing the overall peak to shift to the right. Similarly, for hydrogen bond acceptors as shown in Figure 5D, the peak distribution of the generated molecules was also shifted to the right. This was due to the constraints on lipophilicity during the generation process of the DCT model, which resulted in the generated molecules generally having higher lipophilicity, thereby producing more molecules with multiple hydrogen bond acceptors. Additionally, as shown in Figures 5C and 5E, the distribution curves for the Length of SMIELS and Bertz complexity are all closely aligned with those of the original dataset, indicating that the generative model successfully preserved the key features of molecular size and structural complexity found in the training set.

In Figure 5F, it could be observed that although the overall range of the molecules generated by DCT almost overlapped with the training set, the peak shifting to the right indicated that the molecules generated by DCT had higher synthetic accessibility compared to the training dataset. This aligned with the requirements set for the model during training, where constraints on synthetic accessibility were applied to promote the success of drug synthesis.

RDKit was used to evaluate the validity of the 10,000 generated molecules and to remove any that did not pass validation. First, the SMILES strings of the generated molecules were converted into molecular objects. RDKit's SMILES reading module was used to verify the format of the SMILES strings to ensure the correctness and consistency of the molecular descriptor symbols. The "SanitizeMol" function was then applied to standardize and check the validity of the molecular objects, ensuring they conformed to basic chemical rules. The "DetectChemistryProblems" module was used to identify and address any chemical issues present in the molecules.

**Attribute filtering**

Table 3. Performance of MDAM model

| TASK | DATA_SIZE | SCORE |
| --- | --- | --- |
| BBB | 7,801 | 0.865 |
| SOL/LogS | 31,099 | 1.108 |
| LogP | 249,455 | 1.104 |
| QED | 249,455 | 0.928 |
| SAS | 249,455 | 1.062 |

As shown in Table 3, the attribute model screening was trained and tested on five properties: BBB, LogS, LogP, QED and SAS. BBB was treated as a classification task, with the score evaluated using the F1 score, while the other tests were regression tasks, with scores represented as MSE scores. According to the results in Table 3, the screening model achieved high scores in predicting the five desired properties, indicating its reliable ability to screen potential PROTAC molecules.

**Case study in drug discovery**

The Wnt family was identified as a large family of secreted proteins, consisting of 19 human proteins. As one of the representative signaling proteins of the Wnt family, it was widely distributed and played a crucial role in regulating pleiotropic cellular functions (self-renewal, proliferation, differentiation, and motility). The family was highly conserved and rich in cysteine residues. Wnt signaling proteins mediated their signals by binding to cell membrane receptors and/or adjacent cell membrane receptors. The Wnt signaling pathway was divided into the canonical pathway and the non-canonical pathway. Wnt3a, as the most representative signaling protein of the Wnt family, was widely distributed and played a crucial role in regulating pleiotropic cellular functions (including self-renewal, proliferation, differentiation, and motility). The Wnt signaling pathway was a complex protein interaction network, with its functions most commonly observed in embryonic development and cancer, but also participating in the normal physiological processes of adult animals[78]. The Wnt/β-catenin signaling pathway was one of the most conserved pathways in evolution, playing a key role in embryonic development, cell growth, differentiation, polarity formation, neurodevelopment, and carcinogenesis[79-81]. In previous studies, Wnt3a was found to play a crucial role in the early progression of liver cancer[78, 82].

### *S-mol* extraction

*S-mol* extraction was performed on the target protein Wnt3a, and then the Fusion of Optimal FOTF-CPI prediction model was used to compare all *S-mol*s from the Molecular-*S-mol* library. The calculated SSI values were then ranked, and molecular *S-mol*s with high affinity and their corresponding *S-pro*s were selected based on specific criteria. The selected SSI pairs and their corresponding affinities are presented in Table 4.

Table 4. The top 9 SSI pairs selected based on affinity ranking.

| Index | Wnt3a *S-pro* | *S-mol* | interaction |
|---|---|---|---|
| 1 | EGIKIGIQECQHQFRGRRWNCTTVHDSLAIFGPVLDKATRESAFVHAIASAGVAFAVTRSCAEGT | CN[C@@H](C)C(=O)N[C@H](C(=O)N1CCC[C@H]1C(=O)N[C@H](C(=O)OC)C(C1=CC=CC=C1)C1=CC=CC=C1)C1CCCCC1 | 0.972 |
| 2 | CSEDIEFGGMVSREFADARENRPDA | CC(=O)N[C@@H](CC1=CC=CC=C1)C(=O)N[C@H](C(=O)N1C[C@H](O)C[C@H]1C(=O)NCC1=CC=C(C2=C(C)N=CS2)C=C1)C(C)(C)C | 0.963 |
| 3 | EGIKIGIQECQHQFRGRRWNCTTVHDSLAIFGPVLDKATRESAFVHAIASAGVAFAVTRSCAEGT | O=C1CCC(N2C(=O)C3=CC=CC4=CC=CC(=C34)C2=O)C(=O)N1 | 0.884 |
| 4 | RSAMNRHNNEAGRQAIASHMHLKCKCHGLSGSC | CC(=O)N[C@H](C(=O)N1C[C@H](O)C[C@H]1C(=O)N[C@@H](C)C1=CC=C(C(C)(C)C)C=C1)C(C)(C)C | 0.872 |
| 5 | RSAMNRHNNEAGRQAIASHMHLKCKCHGLSGSC | O=C1CCC(N2C(=O)C3=CC=CC=C3C2=O)C(=O)N1 | 0.871 |

| 6 | CSEDIEFGGMVSREFADARENRPDA | O=C1CCC(N2C(=O)C3=CC=CC=C3C2=O)C(=O)N1 | 0.865 |
| 7 | CSEDIEFGGMVSREFADARENRPDA | CN[C@@H](C)C(=O)N[C@H](C(=O)N1CCC[C@H]1C1=NC(C(=O)C2=CC=CC(OC)=C2)=CS1)C1CCCCC1 | 0.863 |
| 8 | CSEDIEFGGMVSREFADARENRPDA | CNC(=O)C[C@@H](NC(=O)[C@@H]1C[C@@H](O)CN1C(=O)[C@@H](NC(=O)C1(F)CC1)C(C)(C)C)C1=CC=C(C2=C(C)N=CS2)C=C1 | 0.862 |
| 9 | EGIKIGIQECQHQFRGRRWNCTTVHDSLAIFGPVLDKATRESAFVHAIASAGVAFAVTRSCAEGT | CN[C@@H](C)C(=O)N[C@H](C(=O)N1CCC[C@H]1C(=O)N[C@H](C(=O)OC)C(C1=CC=CC=C1)C1=CC=CC=C1)C1CCCCC1 | 0.861 |

## Molecules generation and screening

Based on the 9 selected SSI pairs, a total of 10,000 potential PROTAC molecules were generated by the model, with 100 molecules generated per SSI pair. After validation through RDKit, 9,802 molecules passed the initial screening. A total of 372 molecules were selected for wet experimental validation and screening based on criteria including BBB, SOL(10–100 mg/mL), LogP(1–3), and QED(>0.5). Some of the candidate molecules for wet experimental validation are listed in Table 5.

Reverse synthesis analysis was performed on these language model-driven PROTAC candidate molecules, decomposing them into three parts: protein ligand molecule, linker, and E3 ligase ligand, followed by further analysis of each component. A comprehensive evaluation of the synthetic feasibility and cost efficiency of the 372 selected molecules was then conducted, focusing on the simplicity and feasibility of their synthetic routes, whether special reagents or conditions were required, and overall synthesis cost and efficiency. Preference was given to molecules requiring fewer synthetic steps, lower costs, and those generated from high-affinity *S-mol*s. Ultimately, 12 molecules were selected for further validation.

Table 5 provides a detailed description of the entire screening process.

Table 5: Screening process for PROTAC candidate compounds

| Step | Screening criteria | Molecular number |
|---|---|---|
| 1 | Generated a total of 10,000 molecules using the generation model | 10,000 |
| 2 | RDKit validity check | 9,802 |
| 3 | MDAM screening: TOX/BBB/SOL/LogP | 798 |
| 4 | SAS | 737 |
| 5 | QED/RO5 | 372 |
| 6 | Reverse synthesis analysis | 12 |
| 7 | Molecular dynamics simulation of complexes | 3 |
| 8 | Chemical synthesis | 3 |

Before proceeding with molecular synthesis, the generated PROTAC candidate compounds were validated through molecular docking and molecular dynamics simulations of the molecule-protein interaction. Based on the inverse synthesis results, the three molecules with the best affinity were selected from the synthesizable molecules for validation. After molecular docking and molecular dynamics validation, three compounds that exhibited condition-driven behavior patterns were identified during the molecular dynamics validation process.

**Molecular dynamics simulation**

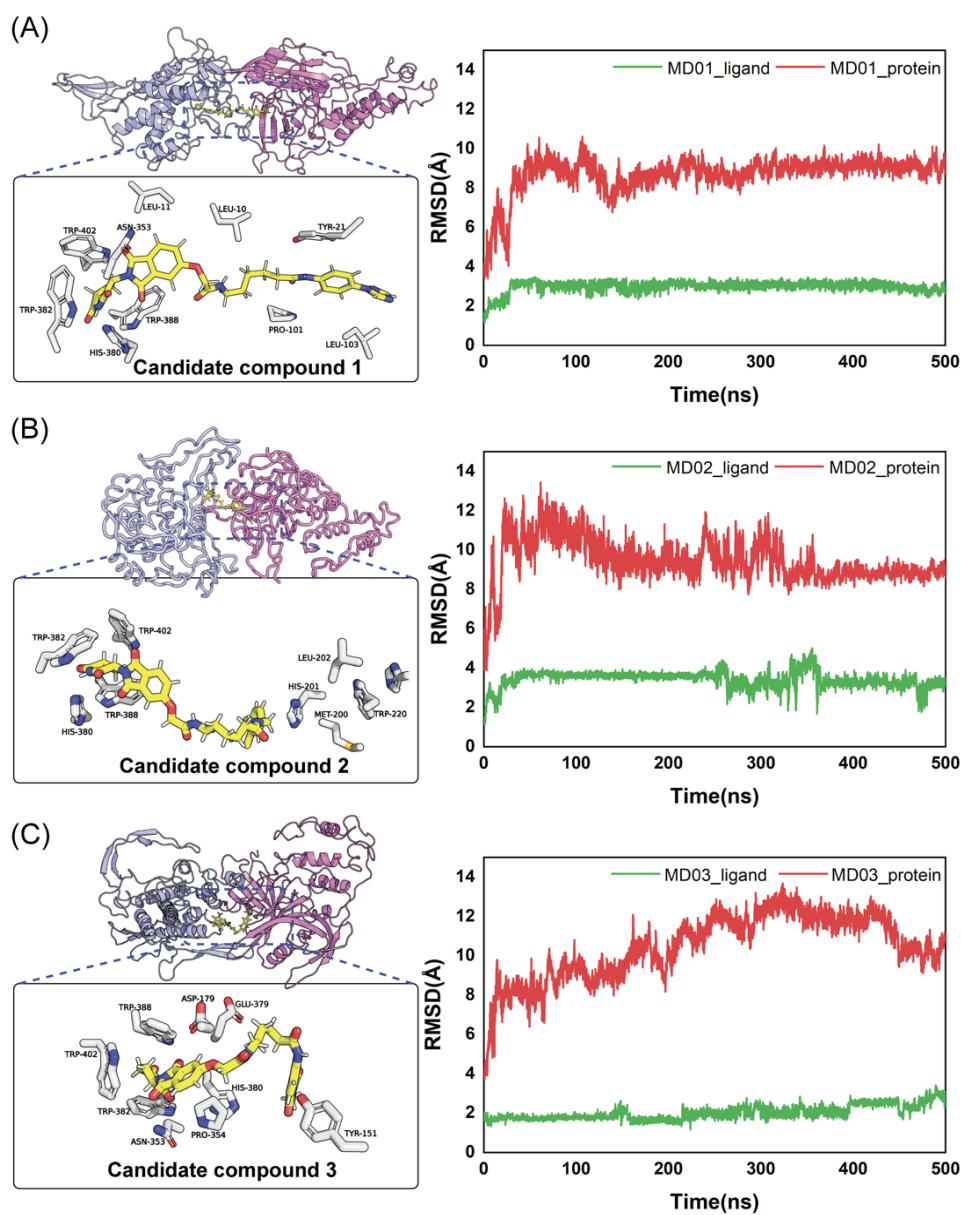

**Figure 6.** Analysis of the Wnt3a and CRBN complex with different Candidate PROTAC ligands in MD simulations. (A) In MD01, RMSD curve, key residue binding energy contributions, and SASA analysis for the first Candidate PROTAC compound (Candidate compound 1). (B) In MD02, RMSD, binding energy, and SASA analysis for the second Candidate PROTAC compound (Candidate compound 2). (C) In MD03, RMSD, binding energy, and SASA analysis for the third Candidate PROTAC compound (Candidate compound 3).

Figure 6A presented the molecular dynamics results for the first Candidate PROTAC compound (Candidate compound 1) with the Wnt3a-CRBN complex, in simulation MD01. This included RMSD, binding energy contributions, and SASA analyses.

In MD01, the RMSD curve stabilized quickly after a minor initial fluctuation, indicating that Candidate compound 1 achieved a stable and lasting binding at the site. This stability suggested good conformational adaptation and affinity.

**Binding Energy Contributions (MM-PBSA):** Residues W382, Y21, and P101 showed substantial negative binding energy, suggesting they played a critical role in stabilizing Candidate compound 1. In particular, W382 likely formed hydrophobic or van der Waals interactions, providing robust anchoring.

The SASA showed a sharp initial decline, suggesting that Candidate compound 1 induced a close-fitting structural arrangement. The significant reduction implied a well-fitted insertion into the binding pocket.

Figure 6B displayed the molecular dynamics results for the second Candidate PROTAC compound (Candidate compound 2) with the Wnt3a-CRBN complex, in simulation MD02. This included RMSD, binding energy contributions, and SASA analyses.

**RMSD Analysis:** In MD02, the RMSD curve showed some fluctuations, indicating a degree of flexibility for Candidate compound 2 within the binding site. However, the fluctuations remained within a stable range, suggesting maintained binding.

**Binding Energy Contributions (MM-PBSA):** Residues W388 and H380 showed significant binding energy contributions, suggesting stable interactions with Candidate compound 2. Despite RMSD fluctuations, these residues' contributions indicated favorable binding potential.

**SASA Analysis:** The SASA curve showed a moderate initial decrease, indicating that Candidate compound 2 achieved a compact structure within the complex, with stability over time.

Figure 6C showed the molecular dynamics results for the third Candidate PROTAC compound (Candidate compound 3) with the Wnt3a-CRBN complex, in simulation MD03. This included RMSD, binding energy contributions, and SASA analyses.

**RMSD Analysis:** In MD03, the RMSD stabilized rapidly with minimal fluctuations, suggesting that Candidate compound 3 maintained a strong, stable interaction at the binding site, indicating high affinity and conformational fit.

**Binding Energy Contributions (MM-PBSA):** Residues N353 and L103 made notable contributions to binding stability, suggesting that these residues supported high affinity for Candidate compound 3 through hydrogen bonds or hydrophobic interactions.

**SASA Analysis:** The SASA curve decreased significantly at first, then stabilized, indicating that Candidate compound 3 induced a compact structure within the complex, supporting enhanced binding stability.

The molecular dynamics simulations of the Wnt3a and CRBN complex with the three Candidate PROTAC compounds revealed distinct interactions and stability across each complex. RMSD analysis demonstrated that all three PROTAC compounds achieved stable interactions within the binding site, with Candidate compound 1 and Candidate compound 3 showing particularly high binding stability. MM-PBSA energy analysis further identified key residues contributing significantly to binding free energy, with residues such as W382, Y21, P101, N353, and L103 playing critical roles in stabilizing the different complexes. Additionally, SASA analysis indicated that the binding of the Candidate PROTAC compounds induced compaction of the complex structure, forming tighter conformations, especially with the first and third compounds.

Overall, the three Candidate PROTAC compounds exhibited favorable binding effects with the Wnt3a and CRBN complex, with Candidate compound 1 and Candidate compound 3 demonstrating superior characteristics in binding stability, energy contributions, and structural compactness. These findings provided valuable structural insights and theoretical support for further optimization and selection of PROTAC compounds in drug design and development. Additional details on the molecular dynamics simulations can be found in Supplementary Information Figure S1.

**Chemical synthesis**

Molecule synthesis: Amide condensation was utilized with mild conditions, high reaction activity, and high yield for splicing. Protein ligand molecules were selected manually based on screening results, choosing rational molecules for synthesis. However, due to synthetic limitations, only two compounds were successfully confirmed by NMR, verifying that their structures were as expected.

The synthesized molecules are shown in Figure 7. Detailed NMR spectra can be found in the supporting information, Figures S2 and S3.

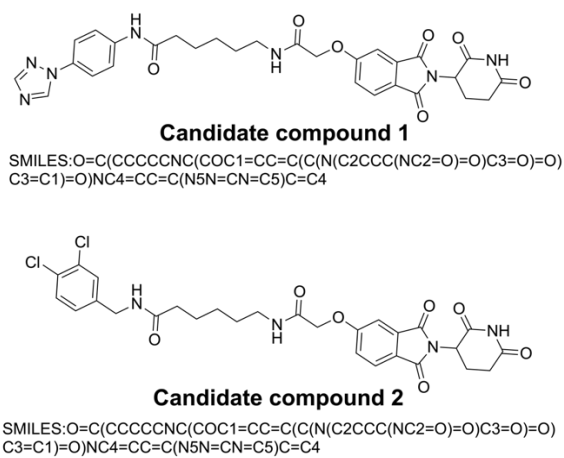

**Figure 7.** Synthesis of Candidate Compounds 1, 2

**In vitro validation**

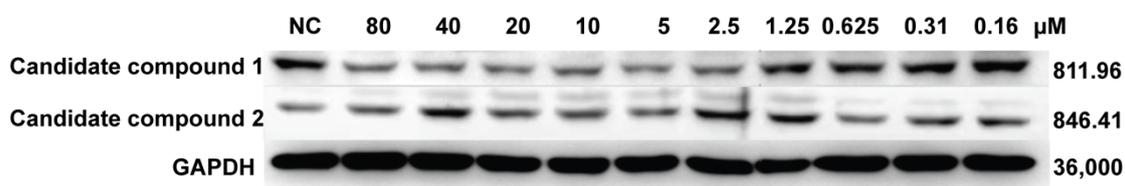

**Figure 8.** Western blot analysis of dose-dependent degradation of target protein by Candidate compounds 1, 2.

As seen in Figure 8, the Western Blot results showed the degradation effects of three candidate PROTAC compounds on Wnt3a protein at various concentrations.

GAPDH bands served as a loading control, ensuring equal protein loading across all samples.

For Candidate Compound 1, a dose-dependent reduction in Wnt3a levels was observed, starting from 0.16 μM and becoming more pronounced up to 5 μM. The gradual decrease in band intensity suggested that Candidate Compound 1 had potential as an effective Wnt3a degrader.

Candidate Compound 2 showed a decrease in Wnt3a levels, with a notable effect beginning at 5 μM. However, the reduction in band intensity was not as substantial or consistent as that of Candidate Compound 1, indicating that while it may have had some effect, its efficacy was comparatively limited at the tested concentrations.

In summary, Candidate Compound 1 emerged as the most promising PROTAC candidate for Wnt3a degradation, whereas Candidate Compounds 2 showed limited degradation potential under the same conditions.

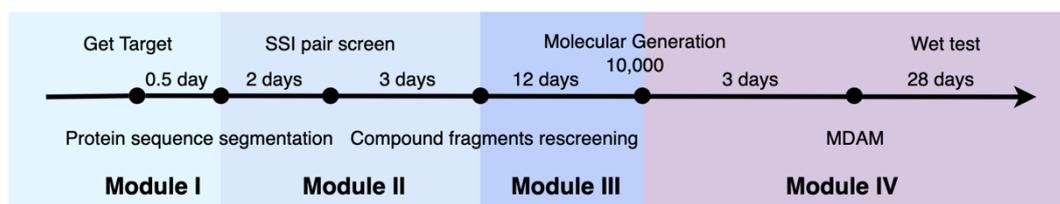

**Figure 9**. Timeline and Workflow of the PROTAC molecule design and validation in LM-PROTAC.

As shown in Figure 9, the entire process of LM-PROTAC, from PROTAC molecule design to experimental validation, was completed in approximately 50 days. Typically, a drug discovery workflow based on artificial intelligence involved significant upfront time for code writing and GPU training/testing, which was not accounted for in the generation process time calculation. This time was not re-spent during the subsequent validation and testing phases. The experiments were conducted in an environment with a 3960X@4.2GHz CPU, A6000*2 with NVLINK, 128GB DRAM, and Ubuntu 20.04 LTS. The process time was the actual runtime recorded in the mentioned software and hardware environment.

**Module I: Target Identification and Data Preprocessing.** This involved obtaining relevant target information, including protein sequence information. In the generation and screening of *S-mol* libraries, the computer played a key role. Segmentation and comprehensive screening of all compounds in the ZINC dataset were performed, which took approximately 0.5 days.

**Module II: Protein-Compound Segment Affinity Screening and Red Screening.** Screening for protein-compound segment affinity took 2 days to identify high SSI pairs. Subsequently, a red screening was performed for the top-ranked SSI pairs, consuming 3 days.

**Module III: Molecules Generation by DCT model** In the subsequent molecular generation work, 10,000 molecules were generated in 12 days.

**Module IV:** The multidimensional attribute screening of these molecules took 3 days, entirely performed automatically by the computer without manual intervention. The selected molecules from this module were synthesized and validated for their activity as PROTAC drugs in wet experiments, which required an additional 28 days for further verification.

The above work constituted a complete solution and was the main workflow for subsequent target-based PROTAC design efforts.

The DCT method proposed in this paper significantly enhanced the efficiency and accuracy of molecular generation by combining *S-mol*s generation techniques with language model-driven molecular generation. The uniqueness of DCT lay in its ability to consider the physicochemical properties of molecules and the specific requirements of the target protein during the generation process, enabling multi-dimensional screening and optimization at the generation stage. This integrated generation and screening mechanism greatly reduced the production of inactive molecules, improved computational efficiency, and enhanced the bioactivity of the molecules. This approach was particularly well-suited for generating complex molecules that required precise targeted design, such as PROTAC molecules.

## Limitations

Based on the experience gained in this study, the efficiency of molecular generation is low, and it is difficult to improve the molecular diversity of PROTAC molecules based on similar compound skeletons. Improvement of molecular diversity is being pursued by expanding the number of molecular segment scaffolds. Simultaneously, efforts are being directed towards enhancing the accuracy of the CPI prediction model to avoid situations where molecules synthesized after screening exhibit lower activity.

## Conclusions

This study presents a de novo language model-driven PROTAC drug generation method that incorporates dual constraints on *S-mol* structure and physicochemical properties based on encoding proteins and compound molecules using the C-Transformer model. Utilizing the FOTF-CPI model to screen *S-mol*s, *S-mol*s were identified with high affinity for the target Wnt3a. Leveraging these screened *S-mol*s and a generative model, a large number of potential candidate compounds that meet the desired criteria were generated. Further refinement of these potential candidates was performed using an attribute selection model to obtain drugs with specific compound properties. The two most promising compounds were synthesized, and additional wet experiments were conducted. The results from the molecular to the cellular level confirmed that one of the three compounds showed inhibitory effects on cancer cells. In comparison to conventional PROTAC generation research, this study employs a molecular splitting approach from the perspective of NLP, enabling the extraction of basic molecular *S-mol*s that align with chemical spatial features from a textual understanding. Additionally, the screened molecules, without further optimization through traditional computational drug discovery methods such as molecular dynamics simulations, demonstrated high activity. This approach reduces manual operations in the overall drug development pipeline, allowing the generation and screening of compliant PROTAC compounds within 18 days. In conclusion, NLP technology provides greater flexibility and automation in PROTAC design. Through segmentation processing, NLP can generate multiple potential candidate molecules in a short time, reducing the need for significant manual intervention. This method significantly enhances the efficiency of PROTAC molecule design, particularly for targets that lack traditional binding pockets. Thus, NLP technology holds great potential for the future development of PROTACs. In the actual molecular generation process, the language model-driven dual-constraint PROTAC generation model can rely entirely on automated computer execution for generation and screening procedures. This program can generate multiple potential compounds meeting the criteria within 20 days without human intervention. These compounds will be further chemically synthesized to validate their inhibitory activity against Wnt3a.

## Author information

### Corresponding author

**Fubo Wang** - Center for Genomic and Personalized Medicine, Guangxi key Laboratory for Genomic and Personalized Medicine, Guangxi Collaborative Innovation Center for Genomic and Personalized Medicine, Guangxi Medical University, Nanning 226001, China; Email: wangfubo@gxmu.edu.cn

**Li Wang** - Research Center for Intelligence Information Technology, Nantong University, Nantong 226001, China; Email: wangli@ntu.edu.cn

### Authors

**Jinsong Shao** - School of Information Science and Technology, Nantong University, Nantong 226001, China; Email: sylershao@gmail.com

**Qineng Gong** - Medical Research Center, Affiliated Hospital 2 of Nantong University and First People's Hospital of Nantong City, Nantong 226001, China

**Zeyu Yin** - School of Information Science and Technology, Nantong University, Nantong 226001, China

**Yu Chen** - School of Information Science and Technology, Nantong University, Nantong 226001, China

**Yajie Hao** - School of Information Science and Technology, Nantong University, Nantong 226001, China

**Lei Zhang** - School of Pharmacy, Nantong University, Nantong 226001, China;

**Linlin Jiang** - School of Pharmacy, Nantong University, Nantong 226001, China;

**Min Yao** - School of Medical, Nantong University, Nantong 226001, China; Email: erbei@ntu.edu.cn

**Jinlong Li** - School of Pharmacy, Nantong University, Nantong 226001, China; Email: jinlongli@ntu.edu.cn

**Funding**

The work was supported by the National Natural Science Foundation of China (No. 32470985), the Science Foundation for Distinguished Young Scholars of Guangxi (2023GXNSFFA026003, F. W.), the Science and Technology Major Project of Guangxi (AA22096030 and AA22096032), the Yongjiang Program of Nanning (2021015, F. W.), the Science Foundation for Distinguished Young Scholars of Guangxi Medical University (F. W.).

**Notes**

The authors declare no competing financial interest.

# References

1. G. R. Hughes, A. P. Dudey, A. M. Hemmings and A. Chantry, Frontiers in PROTACs, *Drug Discovery Today*, 2021, **26**, 2377-2383.
2. L. J. Fulcher, T. Macartney, P. Bozatzi, A. Hornberger, A. Rojas-Fernandez and G. P. Sapkota, An affinity-directed protein missile system for targeted proteolysis, *Open Biol*, 2016, **6**.
3. S. Lim, R. Khoo, K. M. Peh, J. Teo, S. C. Chang, S. Ng, G. L. Beilhartz, R. A. Melnyk, C. W. Johannes, C. J. Brown, D. P. Lane, B. Henry and A. W. Partridge, bioPROTACs as versatile modulators of intracellular therapeutic targets including proliferating cell nuclear antigen (PCNA), *Proc Natl Acad Sci U S A*, 2020, **117**, 5791-5800.
4. C. Mayor-Ruiz, S. Bauer, M. Brand, Z. Kozicka, M. Siklos, H. Imrichova, I. H. Kaltheuner, E. Hahn, K. Seiler, A. Koren, G. Petzold, M. Fellner, C. Bock, A. C. Muller, J. Zuber, M. Geyer, N. H. Thoma, S. Kubicek and G. E. Winter, Rational discovery of molecular glue degraders via scalable chemical profiling, *Nat Chem Biol*, 2020, **16**, 1199-1207.
5. M. Słabicki, Z. Kozicka, G. Petzold, Y.-D. Li, M. Manojkumar, R. D. Bunker, K. A. Donovan, Q. L. Sievers, J. Koeppel and D. Suchyta, The CDK inhibitor CR8 acts as a molecular glue degrader that depletes cyclin K, *Nature*, 2020, **585**, 293-297.
6. K. M. Sakamoto, K. B. Kim, A. Kumagai, F. Mercurio, C. M. Crews and R. J. Deshaies, Protacs: chimeric molecules that target proteins to the Skp1-Cullin-F box complex for ubiquitination and degradation, *Proc Natl Acad Sci U S A*, 2001, **98**, 8554-8559.
7. X. Sun, H. Gao, Y. Yang, M. He, Y. Wu, Y. Song, Y. Tong and Y. Rao, PROTACs: great opportunities for academia and industry, *Signal Transduct Target Ther*, 2019, **4**, 64.
8. D. P. Bondeson, A. Mares, I. E. Smith, E. Ko, S. Campos, A. H. Miah, K. E. Mulholland, N. Routly, D. L. Buckley, J. L. Gustafson, N. Zinn, P. Grandi, S. Shimamura, G. Bergamini, M. Faelth-Savitski, M. Bantscheff, C. Cox, D. A. Gordon, R. R. Willard, J. J. Flanagan, L. N. Casillas, B. J. Votta, W. den Besten, K. Famm, L. Kruidenier, P. S. Carter, J. D. Harling, I. Churcher and C. M. Crews, Catalytic in vivo protein knockdown by small-molecule PROTACs, *Nat Chem Biol*, 2015, **11**, 611-617.
9. M. Zengerle, K. H. Chan and A. Ciulli, Selective Small Molecule Induced Degradation of the BET Bromodomain Protein BRD4, *ACS Chem Biol*, 2015, **10**, 1770-1777.
10. J.-H. Chen and Y. J. Tseng, Different molecular enumeration influences in deep learning: an example using aqueous solubility, *Briefings in Bioinformatics*, 2021, **22**, bbaa092.
11. T. B. Kimber, M. Gagnebin and A. Volkamer, Maxsmi: maximizing molecular property prediction performance with confidence estimation using smiles augmentation and deep learning, *Artificial Intelligence in the Life Sciences*, 2021, **1**, 100014.
12. S. Lim and Y. O. Lee, 2021.
13. M. Hirohara, Y. Saito, Y. Koda, K. Sato and Y. Sakakibara, Convolutional neural network based on SMILES representation of compounds for detecting chemical motif, *BMC bioinformatics*, 2018, **19**, 83-94.

14. C. Li, J. Feng, S. Liu and J. Yao, A novel molecular representation learning for molecular property prediction with a multiple SMILES-based augmentation, *Computational Intelligence and Neuroscience*, 2022, **2022**.
15. B. Fabian, T. Edlich, H. Gaspar, M. Segler, J. Meyers, M. Fiscato and M. Ahmed, Molecular representation learning with language models and domain-relevant auxiliary tasks, *arXiv preprint arXiv:2011.13230*, 2020.
16. R. Liao, Z. Zhao, R. Urtasun and R. S. Zemel, Lanczosnet: Multi-scale deep graph convolutional networks, *arXiv preprint arXiv:1901.01484*, 2019.
17. J. Wang, D. Cao, C. Tang, L. Xu, Q. He, B. Yang, X. Chen, H. Sun and T. Hou, DeepAtomicCharge: a new graph convolutional network-based architecture for accurate prediction of atomic charges, *Briefings in bioinformatics*, 2021, **22**, bbaa183.
18. Z. Xiong, D. Wang, X. Liu, F. Zhong, X. Wan, X. Li, Z. Li, X. Luo, K. Chen and H. Jiang, Pushing the boundaries of molecular representation for drug discovery with the graph attention mechanism, *Journal of medicinal chemistry*, 2019, **63**, 8749-8760.
19. X.-S. Li, X. Liu, L. Lu, X.-S. Hua, Y. Chi and K. Xia, Multiphysical graph neural network (MP-GNN) for COVID-19 drug design, *Briefings in Bioinformatics*, 2022, **23**, bbac231.
20. H. Cho and I. S. Choi, Enhanced Deep-Learning Prediction of Molecular Properties via Augmentation of Bond Topology, *ChemMedChem*, 2019, **14**, 1604-1609.
21. C. Lu, Q. Liu, C. Wang, Z. Huang, P. Lin and L. He, 2019.
22. Y. Liu, L. Wang, M. Liu, X. Zhang, B. Oztekin and S. Ji, Spherical message passing for 3d graph networks, *arXiv preprint arXiv:2102.05013*, 2021.
23. X. Fang, L. Liu, J. Lei, D. He, S. Zhang, J. Zhou, F. Wang, H. Wu and H. Wang, Geometry-enhanced molecular representation learning for property prediction, *Nature Machine Intelligence*, 2022, **4**, 127-134.
24. A. Yoshimori, Prediction of molecular properties using molecular topographic map, *Molecules*, 2021, **26**, 4475.
25. A. B. Tchagang and J. J. Valdés, 2021.
26. J. Iqbal, M. Vogt and J. Bajorath, Learning functional group chemistry from molecular images leads to accurate prediction of activity cliffs, *Artificial Intelligence in the Life Sciences*, 2021, **1**, 100022.
27. M. Langevin, H. Minoux, M. Levesque and M. Bianciotto, Scaffold-constrained molecular generation, *Journal of Chemical Information and Modeling*, 2020, **60**, 5637-5646.
28. Y. Li, J. Hu, Y. Wang, J. Zhou, L. Zhang and Z. Liu, Deepscaffold: a comprehensive tool for scaffold-based de novo drug discovery using deep learning, *Journal of chemical information and modeling*, 2019, **60**, 77-91.
29. W. Jin, R. Barzilay and T. Jaakkola, 2020.
30. M. Podda, D. Bacciu and A. Micheli, 2020.
31. F. Imrie, A. R. Bradley, M. van der Schaar and C. M. Deane, Deep generative models for 3D linker design, *Journal of chemical information and modeling*, 2020, **60**, 1983-1995.
32. H. Green, D. R. Koes and J. D. Durrant, DeepFrag: a deep convolutional neural network for fragment-based lead optimization, *Chemical Science*, 2021, **12**, 8036-8047.

33. M. Xu, L. Yu, Y. Song, C. Shi, S. Ermon and J. Tang, Geodiff: A geometric diffusion model for molecular conformation generation, *arXiv preprint arXiv:2203.02923*, 2022.
34. J. Lim, S. Ryu, J. W. Kim and W. Y. Kim, Molecular generative model based on conditional variational autoencoder for de novo molecular design, *Journal of cheminformatics*, 2018, **10**, 1-9.
35. S. Kang and K. Cho, Conditional molecular design with deep generative models, *Journal of chemical information and modeling*, 2018, **59**, 43-52.
36. S. H. Hong, S. Ryu, J. Lim and W. Y. Kim, Molecular generative model based on an adversarially regularized autoencoder, *Journal of chemical information and modeling*, 2019, **60**, 29-36.
37. W. Jin, R. Barzilay and T. Jaakkola, 2018.
38. B. Samanta, A. De, G. Jana, V. Gómez, P. K. Chattaraj, N. Ganguly and M. Gomez-Rodriguez, Nevae: A deep generative model for molecular graphs, *Journal of machine learning research*, 2020, **21**, 1-33.
39. C. Shi, M. Xu, Z. Zhu, W. Zhang, M. Zhang and J. Tang, Graphaf: a flow-based autoregressive model for molecular graph generation, *arXiv preprint arXiv:2001.09382*, 2020.
40. C. Zang and F. Wang, 2020.
41. RU2745445-C1.
42. A. Nigam, P. Friederich, M. Krenn and A. Aspuru-Guzik, Augmenting genetic algorithms with deep neural networks for exploring the chemical space, *arXiv preprint arXiv:1909.11655*, 2019.
43. V. Bagal, R. Aggarwal, P. Vinod and U. D. Priyakumar, LigGPT: Molecular Generation using a Transformer-Decoder Model, *chemrxiv*, 2021.
44. D. Grechishnikova, Transformer neural network for protein-specific de novo drug generation as a machine translation problem, *Scientific reports*, 2021, **11**, 1-13.
45. N. S. Keskar, B. McCann, L. R. Varshney, C. Xiong and R. Socher, Ctrl: A conditional transformer language model for controllable generation, *arXiv preprint arXiv:1909.05858*, 2019.
46. J. Wang, C.-Y. Hsieh, M. Wang, X. Wang, Z. Wu, D. Jiang, B. Liao, X. Zhang, B. Yang and Q. He, Multi-constraint molecular generation based on conditional transformer, knowledge distillation and reinforcement learning, *Nature Machine Intelligence*, 2021, **3**, 914-922.
47. J. Koehler Leman and G. Künze, Recent advances in NMR protein structure prediction with ROSETTA, *International Journal of Molecular Sciences*, 2023, **24**, 7835.
48. R. Krishna, J. Wang, W. Ahern, P. Sturmfels, P. Venkatesh, I. Kalvet, G. R. Lee, F. S. Morey-Burrows, I. Anishchenko and I. R. Humphreys, Generalized biomolecular modeling and design with RoseTTAFold All-Atom, *Science*, 2024, **384**, eadl2528.
49. J. K. Leman, B. D. Weitzner, S. M. Lewis, J. Adolf-Bryfogle, N. Alam, R. F. Alford, M. Aprahamian, D. Baker, K. A. Barlow and P. Barth, Macromolecular modeling and design in Rosetta: recent methods and frameworks, *Nature methods*, 2020, **17**, 665-680.
50. G. M. Morris, R. Huey, W. Lindstrom, M. F. Sanner, R. K. Belew, D. S. Goodsell and A. J. Olson, AutoDock4 and AutoDockTools4: Automated docking with selective receptor flexibility, *Journal of computational chemistry*, 2009, **30**, 2785-2791.

51. G. Corso, H. Stärk, B. Jing, R. Barzilay and T. Jaakkola, Diffdock: Diffusion steps, twists, and turns for molecular docking, *arXiv preprint arXiv:2210.01776*, 2022.
52. L. G. Ferreira, R. N. Dos Santos, G. Oliva and A. D. Andricopulo, Molecular docking and structure-based drug design strategies, *Molecules*, 2015, **20**, 13384-13421.
53. G. Schneider and U. Fechner, Computer-based de novo design of drug-like molecules, *Nature Reviews Drug Discovery*, 2005, **4**, 649-663.
54. J. Jumper, R. Evans, A. Pritzel, T. Green, M. Figurnov, O. Ronneberger, K. Tunyasuvunakool, R. Bates, A. Žídek and A. Potapenko, Highly accurate protein structure prediction with AlphaFold, *nature*, 2021, **596**, 583-589.
55. L. Pinzi and G. Rastelli, Molecular docking: shifting paradigms in drug discovery, *International journal of molecular sciences*, 2019, **20**, 4331.
56. T. Sterling and J. J. Irwin, ZINC 15--Ligand Discovery for Everyone, *J Chem Inf Model*, 2015, **55**, 2324-2337.
57. M. K. Gilson, T. Liu, M. Baitaluk, G. Nicola, L. Hwang and J. Chong, BindingDB in 2015: a public database for medicinal chemistry, computational chemistry and systems pharmacology, *Nucleic acids research*, 2016, **44**, D1045-D1053.
58. M. I. Davis, J. P. Hunt, S. Herrgard, P. Ciceri, L. M. Wodicka, G. Pallares, M. Hocker, D. K. Treiber and P. P. Zarrinkar, Comprehensive analysis of kinase inhibitor selectivity, *Nature biotechnology*, 2011, **29**, 1046-1051.
59. M. M. Mysinger, M. Carchia, J. J. Irwin and B. K. Shoichet, Directory of useful decoys, enhanced (DUD-E): better ligands and decoys for better benchmarking, *Journal of medicinal chemistry*, 2012, **55**, 6582-6594.
60. Z. Yin, Y. Chen, Y. Hao, S. Pandiyan, J. Shao and L. Wang, FOTF-CPI: A compound-protein interaction prediction transformer based on the fusion of optimal transport fragments, *iScience*, 2023.
61. T. T. Tanimoto, Elementary mathematical theory of classification and prediction, 1958.
62. N. Brown, M. Fiscato, M. H. S. Segler and A. C. Vaucher, GuacaMol: Benchmarking Models for de Novo Molecular Design, *Journal of Chemical Information and Modeling*, 2019, **59**, 1096-1108.
63. D. Rogers and M. Hahn, Extended-connectivity fingerprints, *Journal of chemical information and modeling*, 2010, **50**, 742-754.
64. K. Preuer, P. Renz, T. Unterthiner, S. Hochreiter and G. Klambauer, Frechet ChemNet Distance: A Metric for Generative Models for Molecules in Drug Discovery, *Journal of Chemical Information and Modeling*, 2018, **58**, 1736-1741.
65. P. Ertl and A. Schuffenhauer, Estimation of synthetic accessibility score of drug-like molecules based on molecular complexity and fragment contributions, *Journal of cheminformatics*, 2009, **1**, 1-11.
66. G. R. Bickerton, G. V. Paolini, J. Besnard, S. Muresan and A. L. Hopkins, Quantifying the chemical beauty of drugs, *Nature chemistry*, 2012, **4**, 90-98.
67. H. Öztürk, A. Özgür and E. Ozkirimli, DeepDTA: deep drug–target binding affinity prediction, *Bioinformatics*, 2018, **34**, i821-i829.

68. L. Chen, X. Tan, D. Wang, F. Zhong, X. Liu, T. Yang, X. Luo, K. Chen, H. Jiang and M. Zheng, TransformerCPI: improving compound–protein interaction prediction by sequence-based deep learning with self-attention mechanism and label reversal experiments, *Bioinformatics*, 2020, **36**, 4406-4414.
69. K. Huang, C. Xiao, L. M. Glass and J. Sun, MolTrans: molecular interaction transformer for drug–target interaction prediction, *Bioinformatics*, 2021, **37**, 830-836.
70. M. Wen, Z. Zhang, S. Niu, H. Sha, R. Yang, Y. Yun and H. Lu, Deep-learning-based drug–target interaction prediction, *Journal of proteome research*, 2017, **16**, 1401-1409.
71. Z. Cheng, Q. Zhao, Y. Li and J. Wang, IIFDTI: predicting drug–target interactions through interactive and independent features based on attention mechanism, *Bioinformatics*, 2022, **38**, 4153-4161.
72. A. Ghanbarpour and M. A. Lill, Seq2mol: Automatic design of de novo molecules conditioned by the target protein sequences through deep neural networks, *arXiv preprint arXiv:2010.15900*, 2020.
73. A. Kadurin, S. Nikolenko, K. Khrabrov, A. Aliper and A. Zhavoronkov, druGAN: an advanced generative adversarial autoencoder model for de novo generation of new molecules with desired molecular properties in silico, *Molecular pharmaceutics*, 2017, **14**, 3098-3104.
74. C. Ma and X. Zhang, 2021.
75. G. L. Guimaraes, B. Sanchez-Lengeling, C. Outeiral, P. L. C. Farias and A. Aspuru-Guzik, Objective-reinforced generative adversarial networks (organ) for sequence generation models, *arXiv preprint arXiv:1705.10843*, 2017.
76. V. Bagal, R. Aggarwal, P. Vinod and U. D. Priyakumar, MolGPT: molecular generation using a transformer-decoder model, *Journal of Chemical Information and Modeling*, 2021, **62**, 2064-2076.
77. D. Polykovskiy, A. Zhebrak, B. Sanchez-Lengeling, S. Golovanov, O. Tatanov, S. Belyaev, R. Kurbanov, A. Artamonov, V. Aladinskiy and M. Veselov, Molecular sets (MOSES): a benchmarking platform for molecular generation models, *Frontiers in pharmacology*, 2020, **11**, 565644.
78. W. Zheng, M. Yao, M. Fang, L. Pan, L. Wang, J. Yang, Z. Dong and D. Yao, Oncogenic Wnt3aWnt3a: A Candidate Specific Marker and Novel Molecular Target for Hepatocellular Carcinoma, *Journal of Cancer*, 2019, **10**, 5862-5873.
79. M. Pashirzad, H. Fiuji, M. Khazei, M. Moradi-Binabaj, M. Ryzhikov, M. Shabani, A. Avan and S. M. Hassanian, Role of Wnt3aWnt3a in the pathogenesis of cancer, current status and prospective, *Molecular Biology Reports*, 2019, **46**, 5609-5616.
80. C.-F. Bowin, A. Inoue and G. Schulte, WNT-3A-induced β-catenin signaling does not require signaling through heterotrimeric G proteins, *Journal of Biological Chemistry*, 2019, **294**, 11677-11684.
81. S. He, Y. Lu, X. Liu, X. Huang, E. T. Keller, C.-N. Qian and J. Zhang, Wnt3aWnt3a: functions and implications in cancer, *Chinese Journal of Cancer*, 2015, **34**, 50.
82. L. Pan, M. Yao, W. Zheng, J. Gu, X. Yang, L. Qiu, Y. Cai, W. Wu and D. Yao, Abnormality of Wnt3aWnt3a expression as novel specific biomarker for diagnosis and differentiation of hepatocellular carcinoma, *Tumor Biology*, 2016, **37**, 5561-5568.